\documentclass[conference]{IEEEtran}
\IEEEoverridecommandlockouts
\usepackage{cite}
\usepackage{amsmath,amssymb,amsfonts}
\usepackage{algorithmic}
\usepackage{graphicx}
\usepackage{textcomp}
\usepackage{xcolor}
\usepackage{booktabs}
\usepackage{multirow}
\usepackage{subcaption}
\usepackage{makecell}
\usepackage{textcomp}
\usepackage{hyperref}
\usepackage{amsmath}
\usepackage{colortbl}

\newtheorem{example}{Example}
\newtheorem{definition}{Definition}
\definecolor{Gray}{gray}{0.9}

\usepackage[linesnumbered,ruled,vlined]{algorithm2e}
\usepackage{algorithmic}
\SetAlFnt{\footnotesize}
\SetKw{KwDownTo}{downto}
\SetKw{KwTo}{to}
\SetKw{KwTrue}{true}
\SetKw{KwFalse}{false}
\SetKwInOut{Input}{Input}
\SetKwInOut{Output}{Output}
\SetKw{KwAnd}{and}
\SetKw{KwBreak}{break}

\newcommand{\thename}{\texttt{LOGSAFE}}
\newcommand{\update}[1]{\textcolor{black}{#1}}

\newcommand{\best}[1]{\textcolor{black}{\textbf{\underline{#1}}}}
\newcommand{\second}[1]{\textcolor{black}{\textit{\underline{#1}}}}
\newcommand{\smbf}[1]{\noindent\textbf{#1}}

\usepackage{xcolor}

\usepackage{pifont}

\def\BibTeX{{\rm B\kern-.05em{\sc i\kern-.025em b}\kern-.08em
    T\kern-.1667em\lower.7ex\hbox{E}\kern-.125emX}}

\makeatletter \newcommand{\linebreakand}{\end{@IEEEauthorhalign} \hfill\mbox{}\par \mbox{}\hfill\begin{@IEEEauthorhalign} } \makeatother

\begin{document}

\title{LOGSAFE: Logic-Guided Verification for Trustworthy Federated Time-Series Learning}

\author{
\IEEEauthorblockN{Dung Thuy Nguyen}
\IEEEauthorblockA{\textit{Department of Computer Science} \\
\textit{Vanderbilt University}\\
Nashville, TN, USA}
\and
\IEEEauthorblockN{Ziyan An}
\IEEEauthorblockA{\textit{Department of Computer Science} \\
\textit{Vanderbilt University}\\
Nashville, TN, USA}
\and
\IEEEauthorblockN{Taylor T Johnson}
\IEEEauthorblockA{\textit{Department of Computer Science} \\
\textit{Vanderbilt University}\\
Nashville, TN, USA}
\linebreakand
\IEEEauthorblockN{Meiyi Ma}
\IEEEauthorblockA{\textit{Department of Computer Science} \\
\textit{Vanderbilt University}\\
Nashville, TN, USA}
\and
\IEEEauthorblockN{Kevin Leach}
\IEEEauthorblockA{\textit{Department of Computer Science} \\
\textit{Vanderbilt University}\\
Nashville, TN, USA}
\hfill
}

\author{
\IEEEauthorblockN{\textbf{Dung Thuy Nguyen, Ziyan An, Taylor T. Johnson, Meiyi Ma, Kevin Leach}}
\IEEEauthorblockA{
\textit{Department of Computer Science, Vanderbilt University, Nashville TN}\\
\texttt{\{dung.t.nguyen, ziyan.an, taylor.johnson, meiyi.ma, kevin.leach\}@vanderbilt.edu}
}
}

\maketitle

\begin{abstract}
Federated Learning (FL) offers a promising solution to the privacy challenges associated with centralized Machine Learning (ML) by enabling decentralized, collaborative training across distributed agents and devices commonly found in cyber-physical systems (CPS), such as autonomous vehicle controllers, distributed sensing platforms, and industrial control systems. However, FL remains vulnerable to poisoning attacks, where adversarial clients manipulate local data or model updates to degrade global model performance or induce unsafe behaviors in CPS that depend on these models for sensing, control, and actuation. Although substantial progress has been made in developing defenses against poisoning in FL, existing robust methods are predominantly evaluated on computer vision tasks and fall short of addressing the unique characteristics of time series data that arise naturally in CPS, including temporal dependencies, heterogeneity across devices, and large-scale deployments with potentially many compromised participants.

In this paper, we present \thename{}, a defense mechanism designed to mitigate poisoning attacks in Federated Time Series (FTS), even under heterogeneous client behaviors and high adversarial participation, conditions frequently encountered in large CPS. Unlike traditional model-centric defenses that assess update similarity, \thename{} leverages logical reasoning to evaluate client reliability by aligning their predictions with global time series patterns. Our approach extracts client-specific logical reasoning properties, hierarchically infers global temporal specifications, and verifies each client against these properties to identify and exclude malicious participants during aggregation. Experimental results on two FTS datasets show that \thename{} consistently outperforms existing baselines; in the best case, it reduces prediction error by 93.27 percent relative to the second best method. Our code is available at \url{https://github.com/judydnguyen/LOGSAFE-Robust-FTS}.
\end{abstract}

\begin{IEEEkeywords}
Federated Learning, Poisoning Attacks, Signal Temporal Logic, Time Series.
\end{IEEEkeywords}

\section{Introduction}
\label{sec:intro}

Federated learning (FL) has emerged as a promising paradigm for leveraging data and computing resources from multiple clients to train a shared model under the orchestration of a central server~\cite{mcmahan2017communication}.
In FL, clients use their data to train the model locally and iteratively share the local updates with the server, which then combines the contributions of the participating clients to generate a global update.
Recently, FL has been demonstrated to be efficient for time-series-related tasks~\cite{perifanis2023federated,an2024formal,tripathy2024hybrid,chen2025federated}, and more recently in CPS-related domains~\cite{marfo2025federated,quan2025federated,truong2022light}, enabling secure knowledge sharing across distributed systems while preserving data privacy.
Although FL has many notable characteristics and has been successful in many applications~\cite{wen2023survey,kairouz2021advances,nguyen2024backdoor},
recent studies indicate that FL is fundamentally susceptible to adversarial attacks in which malicious clients manipulate the local training process to contaminate the global model~\cite{pmlr-v108-bagdasaryan20a,nguyen2024backdoor}.
In the context of AI-enabled CPS, such poisoning attacks can translate into unsafe or unstable system behaviors~\cite{figueroa2022adversarial,ibrahum2024deep}.
Attacks against such systems can be broadly classified into \emph{untargeted} and \emph{targeted} attacks~\cite{chakraborty2021survey,rodriguez2023survey,nguyen2024backdoor}.
The former aims to deteriorate the performance of the global model on all test samples~\cite{krum,gaussian_MP};
while the latter focuses on causing the model to generate false predictions following specific objectives of the adversaries~\cite{nguyen2024backdoor,pmlr-v108-bagdasaryan20a}.\\

\begin{figure}[t!]
    \centering
    \includegraphics[width=0.95\linewidth]{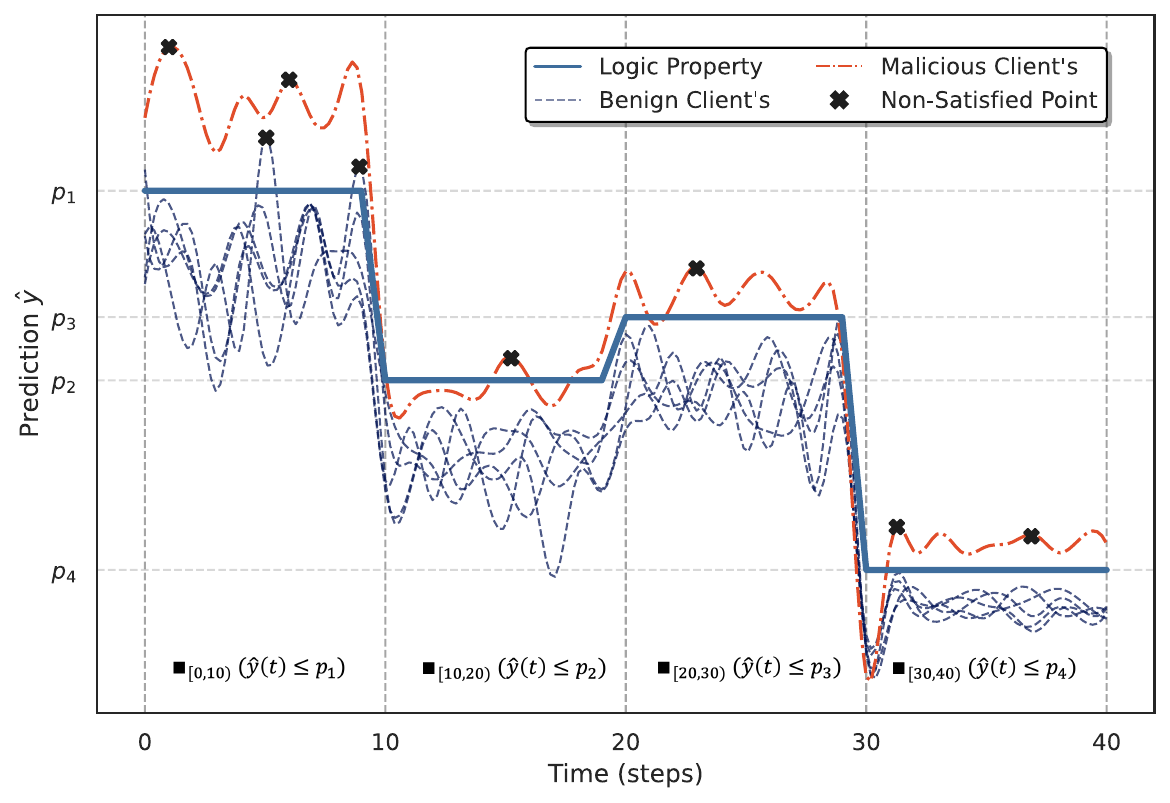}
\caption{Illustration of logical verification given by benign and malicious clients' 
predictions on the FHWA traffic dataset. Our approach learns global properties like 
$\hat{y}(t) \le p_1$ for $t \in (0,10]$, where $\hat{y}(t)$ denotes the predicted traffic 
flow on a highway segment. 
The predictions made by malicious clients often 
violate these properties (marked with $\times$).}

    \label{fig:malicious-behaviors}
    \vspace{-0.5cm}
\end{figure}

\smbf{Motivation.} Prior work has attempted to curtail existing threats in FL, which we broadly classify into two directions: (1) robust FL aggregation~\cite{rfa,rlr,foolsgold} and (2) anomaly model detection~\cite{krum,flame,Rieger2022DeepSightMB,zhang2022fldetector}. %
The former aims to optimize the aggregation function to limit the effects of polluted updates caused by attackers, whereas the latter attempts to identify and remove malicious updates. %
The main drawback of these methods is that polluted updates remain in the global model, reducing the model's precision while leaving the attack impact unmitigated~\cite{nguyen2024backdoor}. 
On the other hand, most methods for identifying malicious clients proposed so far follow the majority-based paradigm in that they assume benign local model updates are aligned and the majority compared to the malicious ones; thus, polluted updates are supposed to be outliers in the distribution of all updates.
In practice, nearly all existing defenses have been designed and evaluated in computer vision contexts, such as MNIST, CIFAR-10, or ImageNet~\cite{krum,trimmed_mean,rfa,foolsgold,flame,zhang2022fldetector,nguyen2024backdoor}, where data modalities are spatial and static. However, these assumptions do not hold in domains such as time-series modeling or CPS, where the data are temporally correlated, non-stationary, and often highly heterogeneous across clients.
Our initial investigation confirms that existing defenses are ineffective in federated time-series (FTS) settings, where temporal heterogeneity and dynamic patterns lead benign updates to diverge substantially. This behavior is much more complex than in image classification tasks, where local models typically converge toward similar feature prototypes for each class, making poisoned updates stand out as outliers. In FTS, by contrast, the lack of such alignment causes malicious updates to blend in with benign ones rather than appear as outliers.
Last but not least, existing defenses lack formal verification and explainability for why a client is considered malicious, making it difficult to audit decisions and undermining trust in safety-critical CPS deployments.
To this end, these limitations become especially critical in CPS, where time-series signals, including network measurements and traffic flows, play a direct role in autonomous decision-making and ensuring operational safety.

\smbf{Our solution.}  To fill this gap, we present
\thename{}, a defense mechanism that mitigates poisoning attacks against Federated Time Series (FTS) under especially challenging scenarios---that is, in the presence of non-IID client data and a large number
of adversarial clients. 
Motivated by recent work demonstrating that STL (Signal Temporal Logic) is highly effective for neural networks on time-series tasks~\cite{ma2020stlnet,an2024formal}, as it provides a formal and interpretable language to specify temporal constraints and domain knowledge that standard neural networks do not inherently capture, we design \thename{} to dynamically mine and verify client-specific STL properties during training, making our method practical across diverse datasets. In our setting, an STL property is a temporal specification over signals, for example, a requirement that whenever the predicted signal exceeds a safety threshold, it must return below that threshold within a fixed number of time steps. These specifications both reflect how temporal patterns are learned from the training data and provide an interpretable description of the model’s behavior over time.
Our approach is orthogonal to existing model-centric defenses that detect or downweight clients based on the similarity of their local updates to one another or to a trusted reference model. Instead, \thename{} evaluates clients based on the satisfaction of STL properties over their predictions, focusing on temporal behavior rather than parameter space. %
Instead, we use STL property mining and verification to study and evaluate each client model's dynamic behaviors, capturing 
traces of predictions on the given time series.
The intuition behind our approach is that, after rounds of training, benign clients naturally reach a consensus of their models' predictions, following close reasoning patterns and sharing STL properties. 
Meanwhile, in the time series context, poisoning attacks disrupt the global model's behavior, leading to abnormal logical patterns in the resulting predictions (e.g., extreme values at specific time steps).
These deviations cause the corresponding logical specifications to diverge from those of the aggregated benign models (illustrated in~\autoref{fig:malicious-behaviors}).\\
First, we extract \textit{local logical reasoning properties} — client-specific STL formulas that summarize recurring temporal patterns in the model’s predictions over time (e.g., in a traffic setting, whenever vehicle density rises above a threshold, the predicted speed drops below a safe limit within a few time steps).
This is conceptually related to dynamic invariant generation in software testing (e.g., DIG~\cite{nguyen2014dig}), but differs in that we mine STL temporal properties over prediction traces, rather than exact algebraic invariants over program executions.
Then, we use unsupervised clustering to group clients based on the parameters of their extracted STL properties (e.g., thresholds, time bounds, and temporal operators).
This enables us to infer \textit{global reasoning properties} that summarize system-wide behavior from the clustered local properties in a hierarchical manner. Instead of forcing all clients to share a predefined specification, we first cluster them based on similar STL property parameters, and then perform aggregation both within and across these clusters. Using unsupervised clustering, which is commonly used to handle heterogeneous client populations in FL~\cite{an2024formal,ghosh2020efficient,caldarola2021cluster}, helps mitigate data heterogeneity (e.g., from differences in hardware, sensors, or control algorithms) while enabling the extraction of coherent global reasoning templates.
These aggregated properties are used to assess the consistency and trustworthiness of client contributions by identifying deviations from expected logical behaviors. 
Our empirical evaluation validates the improved effectiveness of our approach compared to existing defenses that mitigate poisoning attacks by reducing prediction error toward the vanilla case. 

\smbf{Contributions.} We make the following contributions.
(1) We present \thename{} --- a 
 poisoning-resistant defense for Federated Time Series applicable to AI-enabled CPSs. \thename{} identifies and eliminates suspicious clients that distort the global model using logical reasoning property inference and verification, and is, to the best of our knowledge, the first method to thoroughly address both targeted and untargeted poisoning attacks in FTS—even under high ratios of compromised clients and sophisticated attack strategies.
    (2) We systematically evaluate existing robust FL defenses in the context of federated time-series learning (FTS), identify their limitations when adapted to the time-series domain, and demonstrate that formal verification can be leveraged effectively to defend against adversarial attacks in FL.
    (3) We provide in-depth studies on multiple datasets, FL settings, and attack scenarios to demonstrate the superiority of~\thename{} over state-of-the-art defense techniques, highlighting its potential as a step toward safe and verifiable learning in AI-enabled CPS.

\section{Related Works}
\label{sec:related}

\subsection{Poisoning Attacks in FL}
Unlike traditional centralized learning, where data is aggregated at a single location, FL operates in a decentralized setting where data stays on client devices, and only model updates are communicated.
This distributed nature makes FL susceptible to various types of \emph{poisoning attacks}~\cite{nair2023robust,kumar2023impact}. 
A poisoning attack in federated learning occurs when an attacker alters the model submitted by a client to the central server during the aggregation process, either directly or indirectly, causing the global model to update incorrectly~\cite{xia2023poisoning}.%

Depending on the goal of a poisoning attack, we classify poisoning attacks into two categories: (1) untargeted poisoning attacks~\cite{jagielski2018manipulating,krum}, and (2) targeted poisoning attacks~\cite{shafahi2018poison,wang2020attack,DBA}. 
Untargeted poisoning attacks aim to induce the learned model to exhibit a high testing error across all testing examples, thereby resulting in a denial-of-service attack.
The Byzantine attack is among the most popular such attacks~\cite{krum,so2020byzantine}. 
In targeted poisoning attacks, the learned model produces attacker-desired predictions for specific test examples, e.g., predicting spam as non-spam and predicting attacker-desired labels for test examples with a particular trojan trigger (these attacks are also known as backdoor/trojan attacks~\cite{nguyen2024backdoor}). 
In the context of a time-series task, the targeted attack can be adding imperceptible noise to the original sample such that the model predicts the incorrect class~\cite{rathore2020untargeted} or manipulating the prediction into the extreme value/specific directions~\cite{muller2020data}. %

To further strengthen the attack, some model poisoning attacks are often combined with data poisoning ones~\cite{nguyen2024backdoor,pmlr-v108-bagdasaryan20a}. 
In such attacks, adversaries intentionally manipulate their local model updates before sending them to the central server.  
In~\cite{adversarial-lens,nguyen2024iba}, the projected gradient descent (PGD) attack is introduced to be more resistant to many defense mechanisms. 
In a PGD attack, the attacker projects their model on a small ball centered around the previous iteration's global model. 
In addition, scaling and constraint-based techniques~\cite{gaussian_MP,Sun2019CanYR} are commonly used to intensify the poisoning effect while stealthily bypassing robust aggregators.

\subsection{Defenses against Poisoning Attacks in FL}
Defending against poisoning attacks in Federated Learning requires novel approaches tailored to the unique characteristics of the FL paradigm. 
Existing defense strategies can be broadly categorized into (1) robust aggregation mechanisms, (2) anomaly detection, and (3) adversarial training.
Robust Aggregation Mechanisms are designed to mitigate the impact of malicious updates during the model aggregation process~\cite{rlr,krum,rfa,trimmed_mean,karimireddy2022byzantinerobust,allouah2023fixing,karimireddy2021learning}. Blanchard et al.~\cite{krum} proposed the Krum algorithm, which selects updates from a majority of clients that are most similar to each other, thereby excluding potentially malicious updates. 
Another approach, Median-based aggregation~\cite{rfa,trimmed_mean}, takes the median of the updates from all clients, which is more resilient to outliers and adversarial manipulations. 
RLR~\cite{rlr} adjusts the global learning rate based on the sign information contained in each update per dimension and combines differential privacy noise. 

The second approach is backdoor detection, which detects the backdoor gradients and filters them before aggregation~\cite{krum,zhang2022fldetector,foolsgold,huang2023multi,mu2024feddmc,flame,raza2022using,nguyen2023fedgrad}. 
To begin, Cao et al.~\cite{cao2020fltrust} introduced FLTrust, a method that establishes a trust score for each client based on the similarity of their updates to a trusted server-side model. 
Updates with low trust scores are either down-weighted or excluded from the aggregation process. 
More recently, Zhang et al.~\cite{zhang2022fldetector} proposed predicting a client's model update at each iteration from historical model updates and flagging a client as malicious if the received model update from the client and the predicted model update are inconsistent across multiple iterations. 
In addition, Nguyen et al.~\cite{flame} developed an approach that combines anomaly detection with clustering, grouping similar updates and filtering out those that deviate significantly from the majority.  
However, the effectiveness of these methods in Federated Time Series (FTS) scenarios remains unclear, as they lack mechanisms to capture the logical reasoning properties and temporal dependencies inherent in time-series data. Existing model-centric defenses overlook these temporal relationships and dynamic patterns, limiting their ability to detect adversarial behavior accurately.
\\
Recent work has explored using formal logic to improve the interpretability and robustness of federated learning systems. 
An et al.~\cite{an2024formal} proposed a formal logic-enabled personalized FL framework that leverages Signal Temporal Logic (STL) to infer and enforce client-specific properties, using logical constraints as a regularization mechanism for personalization. 
In contrast, our approach is verification-oriented: rather than constraining local training, \thename{} mines global STL specifications from collective client behavior and uses their satisfaction levels as trust signals to detect and mitigate malicious updates during training. 
By integrating STL-based specification mining and verification directly into the learning pipeline, \thename{} provides an adaptive defense mechanism tailored to federated time-series learning. 

\section{Background}
\label{sec:background}
\subsection{Federated Time Series (FTS)}

This work focuses on time series forecasting, which involves predicting future values based on historical data. 
Formally, let $\mathbf{X}_{1: L}=\left(\mathbf{x}_1, \cdots, \mathbf{x}_L\right)^{\top} \in \mathbb{R}^{L \times M}$ be a history sequence of $L$ multivariate time series, where for any time step $t$, each row $\mathbf{x}_t=$ $\left(x_{t 1}, \ldots, x_{t M}\right) \in \mathbb{R}^{1 \times M}$ is a multivariate vector consisting of $M$ variables or channels.
Given a history sequence $\mathbf{X}_{1: L}$ with look-back window $L$, the goal of multivariate time series forecasting is to predict a sequence $\mathbf{X}_{L+1: L+\tau}=\left(\mathbf{x}_{L+1}, \ldots, \mathbf{x}_{L+\tau}\right)^{\top} \in \mathbb{R}^{\tau \times M}$ for the future $\tau$ timesteps.
To achieve this in a decentralized and privacy-preserving manner, we consider a federated learning (FL) system where $N$ participants collaborate to build a global model $G_T$ that generalizes well to their future observations. In our FL system, there is a central server $\mathcal{S}$ and a client pool $\mathcal{C} = [C_1, C_2, \ldots, C_N]$, where each client $i$ has a dataset of size $|\mathcal{D}_i| = N_i$. The training procedure comprises $T$ rounds of interaction between the server and the clients, as follows. \textbf{Step 1. }The server broadcasts the current global model $G^{t-1}$ to all participating clients. Note the initial model $G^0$ is randomly initialized. \textbf{Step 2. }Each client $i$ updates the global model $G^{t-1}$ using its local dataset $\mathcal{D}_i$ via mini-batch stochastic gradient descent (SGD) and sends its updated model weights back to the server. \textbf{Step 3. }The server aggregates the local updates $G^{t}_{i}$ from all clients to generate a new global model $G^{t}$, which is then shared with the clients in the next round.
This iterative process continues until the global model achieves satisfactory performance on all clients' data, ensuring it generalizes well across the diverse datasets in the federated setting.

\subsection{Temporal Reasoning Property Inference}

Signal temporal logic (STL)~\cite{maler2004monitoring} is a precise and flexible formalism designed to specify temporal logic properties. Here, we first provide the syntax of STL, as defined below in Definition~\ref{def:stl}. 
\begin{definition}[STL syntax]\label{def:stl}

$$\varphi 
::= 
\mu 
\mid \neg \mu
\mid \varphi_1\wedge \varphi_2
\mid \varphi_1 \vee \varphi_2  \mid \Diamond_{[a,b]}\varphi 
\mid \square_{[a,b]}\varphi
\mid \varphi_1 \mathcal{U}_{[a,b]} \varphi_2
$$

\end{definition}

\noindent We denote the temporal range as $[a, b] \in \mathbb{R}_{\geq 0}$, where $a \leq b$. Additionally, let $\mu: \mathbb{R}^n \rightarrow \{ \top, \bot \}$ be a signal predicate (e.g., $f(x) \geq 0$) on the signal variable $x \in \mathcal{X}$. Moreover, $\varphi$, $\varphi_1$, and $\varphi_2$ are different STL formulas. The symbols $\square$, $\mathbin{\Diamond}$, and $\mathcal{U}$ are temporal logic operators, denoting ``always,'' ``eventually,'' and ``until,'' respectively. The ``always'' property requires the formula $\varphi$ to be satisfied at all \textit{future} time steps from $a$ to $b$. The ``eventually'' property requires the formula $\varphi$ to be true at some future time steps from $a$ to $b$. Lastly, the ``until'' property requires that $\varphi_1$ is true until $\varphi_2$ becomes true.\\
We suggest referring to An et al.~\cite{an2024formal} for eight examples that can be expressed using STL and inferred through specification mining algorithms. However, users are not restricted to these specifications; they can define any specifications that can be parsed into STL formulas and syntax.
The formal description of the logic property inference task is shown in Definition~\ref{def:stlinf} below~\cite{bartocci2022survey}. Additionally, we provide a practical example of STL property inference in Example~\ref{eg:tempstl}. 
\begin{definition}[STL property inference]\label{def:stlinf}
Given an observed fact $x$ and a parameterized STL formula $\varphi(\alpha)$, where parameter $\alpha$ is unknown, the task is to find the correct value of $\alpha$ such that the formula $\varphi$ holds for all instances of $x$. 
\end{definition}
\begin{example}[Example of STL inference]\label{eg:tempstl}
Consider the template STL property $\varphi_m (\alpha) = \square_{[0, 5)}( (x_1 \leq 0.5) \lor (\alpha \leq x_2 \leq 10 ))$. The task is to find a value for the unknown parameter $\alpha$ such that during future time stamps between $0$ and $5$ if the signal variable $x_1$ is less than or equal to $0.5$, the signal variable $x_2$ must be greater than or equal to $\alpha$ and less or equal than $10$. 
\end{example}

\noindent While $\alpha$ can take infinitely many values, only certain ones enhance reasoning during the verification process. The choice of $\alpha$ influences how well a property $\mathcal{X}$ aligns with observations, shaping how accurately the STL property reflects the data. 
Therefore, the STL inference task can be better formulated in Definition~\ref{def:stltight}, which introduces the concept of a tight bound.

\begin{definition}[STL property inference under a tight bound]\label{def:stltight}
Given observed data $\mathcal{X}$ and a templated STL formula $\varphi (\alpha)$, where $\alpha$ is an unknown free parameter, the objective is to determine a value for $\alpha$ that yields a tightly-fitted temporal logic property. The objective is represented by the equation $\rho (\varphi, \mathcal{X}; \alpha) = \epsilon$, where a smaller positive $\epsilon$ indicates a closer alignment with the data. 
\end{definition}
The task presented in Definition~\ref{def:stltight} can be transformed into an equivalent numerical optimization problem, which can be effectively solved using both gradient-free and gradient-based off-the-shelf solvers, as demonstrated below~\cite{jha2017telex}.

\begin{equation*}\label{eq:solvestl}
\begin{gathered}
    \mathop{\text{min}}{|\epsilon|} \; s.t. \; \epsilon=p^\prime-p  \\ 
    \text{where} \; \rho(\varphi(p),\mathcal{X},t) \geq 0 \; \text{and} \; \rho(\varphi(p^\prime),\mathcal{X},t) < 0
\end{gathered}
\end{equation*}
In a templated logic formula $\varphi$, candidate values of free parameters are denoted by $p$ and $p^\prime$ and the timestamp is denoted by $t$. As described in the equation above, the goal is to minimize the value of $|\epsilon|$, which measures the difference between a satisfactory parameter value and an unsatisfactory one.
Since the STL robustness function is non-differentiable at zero, alternatives such as the ``tightness metric'' in Jha et al.~\cite{jha2017telex} can be utilized to address this problem. 
We further provide Example~\ref{eg:tight-bound} to illustrate this definition better.
\begin{example}[Example of tight-bound inference]\label{eg:tight-bound}
Consider the task from Definition~\ref{def:stltight} with the example of finding $\alpha$ that satisfies $\varphi_m(\alpha) = \square_{[0, 5)}( (x_1 \leq 0.5) \lor (\alpha \leq x_2 \leq 10 ))$ for the 5-step sequence $\mathcal{X} = ((0.4, 4), (0.45, 5), $ $(0.55, 6), $ $(0.75, 7), (1.0, 9))$. The goal is to find $p$ to minimize the value of $|\epsilon|$.
\end{example}
We evaluate $p = 5$, $p = 6$, and $p' = 7$.
For $\alpha = 5$ and $\alpha = 6$, the formula is satisfied at every time step. However, for $\alpha = 7$, $\rho(\varphi_m(7), \mathcal{X})$ is negative at $t = 3$, indicating a violation.
To minimize the discrepancy $\epsilon = p' - p$, we seek smaller $\epsilon$. For instance, $\epsilon = 1$ when $p = 6$ and $p' = 7$, compared to $\epsilon = 2$ for $p = 5$ and $p' = 7$. Thus, $\alpha = 6$ provides a tighter fit than $\alpha = 5$, reducing the mismatch between the STL formula and observed data, ensuring better alignment with the logic specification.

\subsection{Threat Model}

\noindent\textbf{Attacker Capabilities}:
We consider an FL system consisting of a central server \(\mathcal{S}\) and a pool of clients \(\mathcal{C}\) with size $N$, where a fraction $\epsilon$ of these clients are malicious, i.e., \(\epsilon = \frac{|\mathcal{C}_p|}{N}\). These malicious clients can operate independently (non-colluded) or be controlled by a single adversary $A$ (colluded). 
The objective of the malicious clients is to compromise the global model \(G_T\) by introducing poisoned time series data or tampering with model updates. The primary goal is to manipulate the poisoned model \(G'_T\) to output the predictions in the poisoning objective that the adversaries expect.
In our study, we assume that malicious clients can render \textit{targeted} or \textit{untargeted} attacks.
In the untargeted case, the objective is to produce predictions with a high testing error indiscriminately across testing examples, which is represented as:
$\max ( \mathcal{L}(G_T(\mathcal{X}), \mathcal{Y}) )$
where \(G_T(\mathcal{X})\) denotes the predictions of the model \(G_T\) on input \(\mathcal{X}\), and $\mathcal{Y}$ represents the true labels.
Meanwhile, in the targeted case, the goal is to minimize the loss between the model's predictions and the adversarial target labels:
$\min (\mathcal{L}(G_T(\mathcal{X}), \hat{Y}))$,
where \(\hat{Y}\) is the adversarial target label.
In addition, we consider malicious clients with \textit{white-box} attack capabilities. This assumption implies that these clients can manipulate their local training data, injecting poisoned time-series data $\mathcal{D}_p$ into the training set $\mathcal{D}'_i$ of each compromised client $i$. They can also control the local training process and modify model updates before sending them to the server for aggregation, which is more challenging for the defender.
    
\noindent\textbf{Defender Capabilities}: 
Following existing FL defenses, we assume the defender, represented by the server, has the following capabilities. The defender can only access local model updates after the local training process and has no prior knowledge of the attack strategies employed by adversaries. Furthermore, the defender can access only a limited portion of unlabeled clean data, rather than the full training set. It is important to note that the assumption of clean data is not unique to our method and has been widely adopted in FL-related research fields~\cite{liu2018trojaning,zhu2023adfl,Rieger2022DeepSightMB}. Additionally, we assume the server is honest and does not engage in privacy-breaching attacks. Finally, we assume the server has sufficient computational resources, which aligns with real-world scenarios~\cite{zhu2023adfl}. The defender's primary objective is to minimize malicious clients' influence on the global model by reducing the empirical prediction error.

\section{Approach}
\label{sec:method}
\begin{figure*}[t!]
    \centering
    \includegraphics[width=0.95\linewidth]{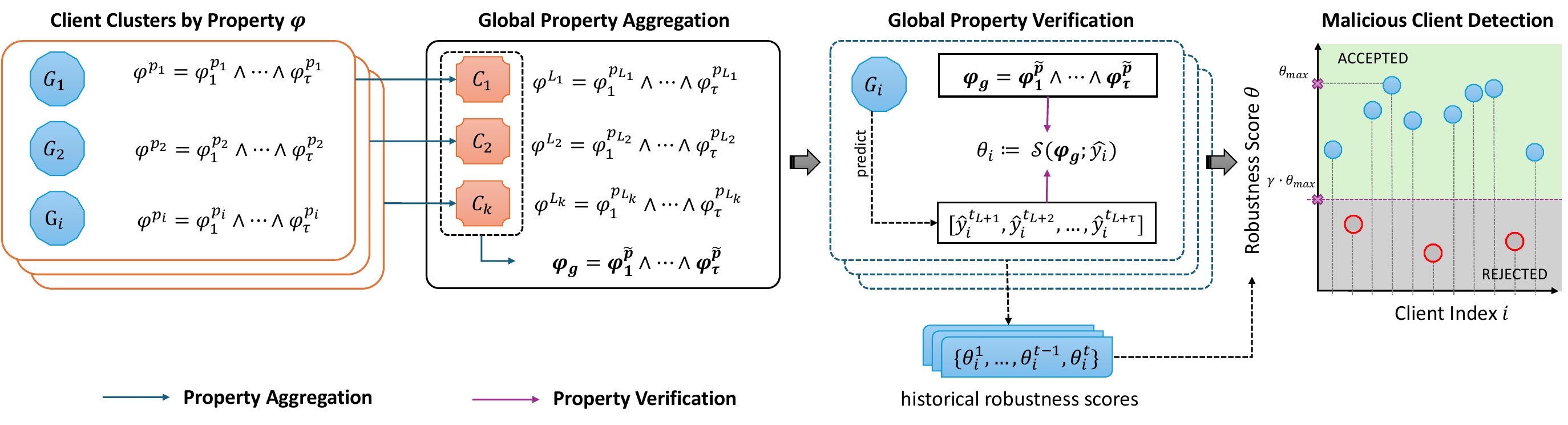}
    \caption{Overview of \thename{}. For each training round, \thename{} first conducts local logic inference to acquire the local reasoning properties; which are then used as a criterion to cluster clients. Global property $\varphi_g$ is calculated by aggregation of clustered properties. Using $\varphi_g$, the server $\mathcal{S}$ verifies the satisfaction scores $\theta$ for each client and uses them to determine the suspicious updates to be removed from the final aggregation.}
    \label{fig:overview}
    \vspace{-0.5cm}
\end{figure*}

Figure~\ref{fig:overview} illustrates the four components that comprise \thename{}: (i) local logic inference; (ii) global logic inference; (iii) global property verification, and (iv) malicious client detection. In this section, we will present in detail how each step is performed. 

\subsection{Local Logic Inference}
\label{method:step1}
\thename{} operates in an FL scenario, where each client completes local training to compute an update that is communicated each training round with an aggregating server. 
During each such communication round, we mine STL specifications from each client after completing its local training.
We employ stochastic gradient descent (SGD)~\cite{an2024formal} as the optimization algorithm to update the models 
as: $G^t_i \leftarrow \operatorname{SGD}\left(G^t, D_i, \eta\right)$, 
where $\eta$ is the learning rate.

Given a local model of the client $C_i$ during round $t$, i.e., $G_i^t$, and $\mathcal{D}_v$ be the unlabeled validation data owned by the server, our method generates the logic property $\varphi (p^i)$ for each participating client based on their prediction $\mathcal{Y}^t_i$ on dataset $\mathcal{D}_v$. 
Our method uses Equation~\ref{eq:solvestl} to infer a logic reasoning property $\varphi$ from the client prediction. 
Recall that any locally inferred logic reasoning property must be satisfied by every data point in the client's prediction. 
Moreover, any STL formula can be represented by its equivalent Disjunctive Normal Form (DNF), which typically has the form of $P \vee Q \vee R \vee \ldots$.
Each clause within a DNF formula consists of variables, literals, or conjunctions (e.g., $P:=p_1 \wedge \neg p_2 \wedge \ldots \wedge p_n$). 
Essentially, the DNF form defines a range of satisfaction where any clause connected with the disjunction operator satisfies the STL formula $\varphi$. 
After this step, each client $C_i$ has a logical property represented $\varphi(p^i):=\varphi_1(p^i) \wedge \varphi_2(p^i) \wedge \ldots \wedge \varphi_n(p^i)$ extracted by logic inference.

\subsection{Global Logic Inference}

Given a set of local logic properties \(\mathcal{P} = \{\varphi(p^1), \varphi(p^2), \ldots, \varphi(p^m)\}\) at the \(t^{th}\) round, the server aggregates them to construct the global property \(P_g\) that all clients should satisfy. The core idea involves multi-level aggregation, where clients with similar properties are first clustered to derive a global property for each cluster. These cluster-level properties are then aggregated to form the final global logic properties.
This approach helps mitigate targeted and untargeted attacks, as malicious clients with properties considerably different from benign ones can be isolated, reducing their impact on the global property \(P_g\).

To categorize samples based on their properties, we use FINCH~\cite{sarfraz2019efficient}, a hierarchical clustering method that identifies groupings in the data based on neighborhood relationships, without requiring hyperparameters or predefined thresholds. 
FINCH is an unsupervised clustering algorithm particularly suitable for scenarios with an uncertain number of clusters, making it ideal for clustering client properties.
Using FINCH, we group clients based on their local properties \(\{P_i\}\) as follows:
\(
\mathcal{P} \xrightarrow{\text{FINCH}} \Gamma_{L_k} = \{\xi_i\}_{i=1}^{K} \wedge r_{i, k} \in \mathbb{R}^{m * K}.
\)
Here, $\{\xi_i\}_{i=1}^{K}$ is a set of clusters, and $\mathbf{r}$ is the cluster assignment metric, where $r_{i, k} \in \mathbb{R}^{m * K}$ and $r_{i, k}=1$ if $i \in k$ else $r_{i, k}=0$.
This technique implicitly clusters clients with similar logical properties, as clients with similar temporal reasoning properties are grouped, while those with differing properties are not combined.
A key advantage of the \thename{} framework is its dynamic assignment of client properties to clusters, allowing the aggregation process to adapt to changes in logic properties over time. 
Next, we discuss how to extract cluster property for each cluster $\xi_i$.

\textbf{Cluster Property Inference.} For a given cluster $\xi_i = \{\varphi(p^j) |\ \forall j \in [1, m] \wedge r^{j,i} = 1\}$, where local property \(\varphi(p^j)\) is from client $j^{th}$ in cluster $\xi_i$, the corresponding cluster property $P_{L_i}$ is computed as:
\begin{equation}
    \label{eqn:cluster-stl}
    P_{L_i} := \varphi(\bar{p}), \quad \bar{p} := \frac{1}{|\xi_i|} \sum_{j=1}^{|\xi_i|} p_j       
\end{equation}
In Equation~\ref{eqn:cluster-stl}, %
\(\bar{p}\) represents the average of the properties inferred by the local clients within the cluster, and \(P_{L_i}\) denotes the cluster property expressed in the same STL (Signal Temporal Logic) syntax as the local models.
The rationale behind this approach is to derive a representative property for the cluster by aggregating the individual properties of the clients. 
By averaging the properties, we create a cluster property that captures the commonalities among clients in the cluster, thus providing a robust and collective representation of their logic properties.

\textbf{Global Property Inference.} Given a set of cluster properties $\{P_{L_i} = \varphi(p^{L_i}) | \forall i \in [1, K] \}$, the global property is then calculated as the median of these parameters from cluster properties, which is:
\begin{equation}
    P_g := \varphi(\tilde{p}), \quad \tilde{p} := Med (p^{L_i}),  \forall i \in [1, K],
\end{equation}

Clustering clients is a well-established approach for handling heterogeneous and non-IID data, where local properties' parameters vary greatly~\cite{caldarola2021cluster,sattler2020clustered}.%
Inspired by this, we construct the global property by averaging cluster properties rather than individual local ones. This ensures the global property better represents each cluster's data, enhancing model generalization within groups and avoiding a uniform approach that may not fit diverse client distributions. By leveraging the median, which is less sensitive to outliers, we reduce the influence of adversarial clients with properties that deviate sharply from benign ones.
This aggregation process aligns to build a global model that is resilient to both targeted and untargeted attacks, as it reduces the impact of malicious clients whose properties deviate dramatically from the benign ones.

\subsection{Global Property Verification}
Given a global property $\varphi(\tilde{p})$, this component assesses the level of satisfaction for each percentage of network predictions, denoted by $\hat{\mathcal{Y}} = (y_{L+1}, \ldots, y_{L+\tau})$, with respect to a specified property $\varphi$. 
Given an input history $\mathcal{X} = (x_1, \ldots, x_L)$, the temporal logic formula $\varphi$ is defined over the resulting predicted trace $\hat{\mathcal{Y}}$, and its satisfaction value quantifies the degree to which $\hat{\mathcal{Y}}$ over the next $\tau$ time steps satisfies $\varphi$ under the condition that $\mathcal{X}$ is provided as the preceding context.
This framework enables us to quantify the degree to which the network's predictions align with the desired property. We formalize the concept of a robustness score $\theta$ in Definition~\ref{def:robustness-score}.

\begin{definition}[Robustness Score $\theta$]\label{def:robustness-score} Given a network's prediction $\hat{\mathcal{Y}}$ $=\left(y_{L+1}, \ldots, y_{L+\tau}\right)$, logical property $\varphi$ and scoring function $\mathcal{S}$, robustness score $\theta$ quantify the degree to which the predicted sequence $\hat{\mathcal{Y}}$ adheres to the specified property $\varphi$. Formally,
\(
    \theta := \mathcal{S}(\varphi, \hat{\mathcal{Y}}) = \{\rho(\varphi, \mathcal{Y}, t) | \ \forall t \in [L+1, L+\tau]\},
\)
where $\rho(\varphi, x, t)$ maps an STL formula $\varphi$, a signal trace $\mathcal{Y}$ at time $t$, to a value. 
\end{definition}
Depending on the chosen scoring function, $\rho$ returns either a binary satisfaction status or a real-valued satisfaction level~\cite{maler2004monitoring,jha2017telex}. We provide a practical example of STL property verification in Example~\ref{eg:verification}.
\begin{example}[Example of STL verification]\label{eg:verification}
Given the STL property $\varphi = \square_{[0, 5)} ((x_1 \geq 0.2)$ $\wedge$ $(x_1 \leq 2.5)$ $\wedge$ $(x_2 \geq 6) \wedge (x_2 \leq 10))$, we evaluated the robustness score $\theta$ using predicted sequences $\hat{\mathcal{Y}} = [(0.4, 4), (0.45, 5), (0.55, 6), (0.75, 7), (1.0, 9)]$. The scoring function $\rho(\varphi, \mathcal{Y}, t)$ calculates the degree of satisfaction at each time step by checking whether both $x_1$ and $x_2$ lie within their specified bounds and returns a boolean value. The final score is $\theta = [\bot, \bot, \top, \top, \top]$, where $\top$ means the property is satisfied, while $\bot$ means the property is violated at that time step.

\end{example}

For each training round, after the robustness score is calculated using Definition~\ref{def:robustness-score}, the server collects a set of scores for each client, denoted as $\theta^t := \{\theta_1^t, \theta_2^t, \ldots, \theta^K_t\}$. To convert the robustness score $\theta$ from qualitative semantics (i.e., boolean values) to numerical values, we compute the fraction of ``True'' signal predicates (i.e., the fraction of $\top$ in $\theta^t$). 
A higher-scoring $\theta$ indicates that a client's predictions align with global properties or the majority of benign clusters, reflecting model reliability and suggesting the client is likely benign. Conversely, lower-scoring $\theta$ highlights discrepancies in the client's model or data, potentially signaling anomalies or malicious intent, warranting further investigation.
Over time, we hypothesize that benign clients increasingly agree on the training objective, leading to more similar models and higher alignment in their predictions on the same validation set. As a result, these clients are expected to achieve higher $\theta$ scores, reflecting consistent adherence to the global property and reinforcing their benign behavior.

To further enhance the reliability of the robustness score, we use historical information to update scores cumulatively. Using historical information has been used in related works~\cite{karimireddy2021learning,foolsgold,karimireddy2022byzantinerobust} and demonstrated its potential improvement. Specifically, the score for client $i$ is updated as $\theta_i = \frac{f_i-1}{f_i} \theta_i + \frac{1}{f_i} \theta_i^t, $
where $\theta_i$ is the cumulative score, $\theta_i^t$ is the current round's score, and $f_i$ represents the number of training rounds client $i$ has participated in up to round $t$. 
By using cumulative scores, the method evaluates each client's overall consistency and performance across multiple rounds, reducing the impact of erratic single-round results. This trustworthy score helps identify and exclude malicious clients from aggregation during each training round.

\subsection{Malicious Client Detection}
To identify and filter out malicious clients, we use a binary mask $\mathcal{M} \in \mathbb{R}^m$, where each element $\mathcal{M}_i$ indicates whether a client $i$ is considered malicious. The mask is defined as follows:
\begin{equation}
\label{eqn:update-mask}
    \mathcal{M}_i = \begin{cases}
    1 & \text{if } \theta_i \geq \gamma \cdot \max{(\theta_j)}, \\
    0 & \text{otherwise}.
    \end{cases}
\end{equation}
Here $\gamma$ is a threshold parameter determining the cutoff for malicious model filtering. Note that the trustworthy scores $\theta$ are normalized before applying Equation~\ref{eqn:update-mask}.
The rationale behind this approach is that models producing predictions with significantly lower robustness scores are more likely to be compromised or adversarial, especially if they are in the minority. The assumption is that the properties of poisoned models, which are often outliers or deviate from the global property, will exhibit lower scores than most benign clients.
The robustness threshold for each training round is determined based on the distribution of the robustness scores across all clients. This threshold helps distinguish between reliable and unreliable client updates. To ensure the integrity of the global model, only updates from non-malicious clients are used for aggregation. The global model $\boldsymbol{w}_t$ at round $t$ is updated as follows:
\begin{equation}
\boldsymbol{G}_t = \mathcal{A}(\boldsymbol{G}_{t-1}; \{\nabla_i \mid \forall i \in [1, m] \text{ and } \mathcal{M}_i = 1\}),
\end{equation}
where $\mathcal{A}$ represents the aggregation function used by the server to combine the updates from the selected non-malicious clients. In this formulation, $\nabla_i$ denotes the model update from client $i$. By excluding updates from clients marked as malicious (i.e., those with $\mathcal{M}_i = 0$), the server can ensure that the global model $\boldsymbol{w}_t$ is not adversely affected by compromised or unreliable client contributions. We hypothesize that~\thename{} enhances the robustness of the global model and helps maintain its performance by focusing on contributions from trustworthy clients.

\section{Experiments}
\label{sec:experiments}
In this section, we empirically evaluate the performance of \thename{} under various poisoning attack scenarios, including both untargeted and targeted attacks. We compare \thename{} against state-of-the-art FL defenses, including Krum/Multi-Krum~\cite{krum}, FoolsGold~\cite{foolsgold}, RFA~\cite{rfa}, ARAGG~\cite{karimireddy2022byzantinerobust}, RLR~\cite{rlr}, FLAME~\cite{flame}, and FLDetector~\cite{zhang2022fldetector}. 
All experiments are conducted on an Amazon EC2 g5.4xlarge instance equipped with one NVIDIA A10G Tensor Core GPU, 16 vCPUs, 64 GB of memory, and 24 GB of dedicated GPU memory.
To thoroughly evaluate the robustness and generalization capability of \thename{}, we organize our study around the following research questions:
\begin{itemize}
    \item \textbf{RQ1:} Can \thename{} robustly defend against both untargeted and targeted poisoning attacks?
    \item \textbf{RQ2:} Does \thename{} generalize across diverse model architectures and FL aggregators?
    \item \textbf{RQ3:} How does \thename{} behave under varying adversarial strengths, including different attack ratios and strategies?
\end{itemize}

\subsection{Experiment Setup}
\noindent\textbf{Dataset descriptions. } 
Following the previous FTS studies~\cite{perifanis2023federated,an2024formal}, we conduct experiments using two time-series datasets, which are LTE Physical Downlink Control Channel (PDCCH) measurements~\cite{perifanis2023federated} and Federal Highway Administration~\cite{hpmsfield}. 
The PDCCH dataset was collected from three different base stations in Barcelona, Spain, which is publicly accessed by Perifanis et al.~\cite{perifanis2023federated}. 
We further separate data from these three stations into 30 sub-clients to simulate the FL environment.
Following the settings by Perifanis et al.~\cite{perifanis2023federated}, the objective is to predict the first five measurements for the next timestep using as input the observations with a window of size $\tau = 10$ per base station. 
The FHWA dataset is a real-world dataset of highway traffic data. 
We obtained a publicly available dataset from the Federal Highway Administration (FHWA 2016) and preprocessed hourly traffic volume from multiple states. After preprocessing and excluding low-quality and missing data, we continue with the final data from 15 states, this setting is followed~\cite{an2024formal}. The learning objective is to predict the traffic volume for the next $\tau = 24$ consecutive hours based on the past traffic volume at a location over the previous five days. 
For these two datasets, we focus on using operational range (ref.~\cite{an2024formal}) for logical reasoning inference and qualitative semantics for verification.

\noindent\textbf{Attack settings. } 
We argue that existing methods typically handle either targeted or untargeted attacks, but not both. However, from a defender’s perspective, it is difficult to anticipate which type of attack an adversary will launch, making it more practical to develop robust FL methods that can effectively handle both.
Therefore, we compare~\thename{} with other defenses under both (i) \textit{untargeted attacks} and (ii) \textit{targeted attacks}.  
Regarding \textit{untargeted attacks}, we consider Gaussian Byzantine attack followed by Fang et al.~\cite{gaussian_MP}, where for a Byzantine client $i$ in iteration $r$, the message to be uploaded is set to follow a Gaussian distribution, i.e., $ \nabla_i^t \sim \mathcal{N}\left(0, \sigma_{\mathrm{G}}^2\right)$. This attack randomly crafts local models, which can mimic certain real-world scenarios where compromised devices might produce erratic or unexpected updates. Regarding \textit{targeted attacks}, we consider the case where malicious clients use a flip attack to manipulate the training data, i.e., data poisoning. Specifically, the adversary aims to disrupt a machine learning model by modifying a subset of data points to their extreme target values~\cite{muller2020data}. This attack is conducted by computing the distance of each target value $\mathcal{Y}_t \in \mathcal{Y}$
from the nearest boundary (i.e., $\mathcal{Y}_{min}, \mathcal{Y}_{max}$) of the feasibility domain, then selecting the points with the largest distances and flipping their target values to the opposite extremes. Under this attack, the victim model shifts the target variable to the other side's extreme value of the feasibility domain\footnote{We reproduce this attack using source code published by Muller et al.~\cite{muller2020data}.}. \\
In FL, to make poisoning attacks stronger and more stealthy against defenses, we evaluate \textit{white-box} attacks, where the adversary can combine data poisoning (as discussed above) with model poisoning techniques. 
Consequently, when evaluating targeted attacks, we examine four strategies: (i) Targeted Attack (without model poisoning), (ii) PGD Attack (Targeted Attack combined with PGD), (iii) Constrain-and-Scale Attack (Targeted Attack combined with Constrain-and-Scale), and (iv) Model Replacement Attack (Targeted Attack combined with Model Replacement)~\cite{pmlr-v108-bagdasaryan20a,nguyen2024backdoor}. 

\noindent\textbf{Defense baselines. }We implement seven representative FL defenses for byzantine attacks and targeted attacks, including Krum/Multi-Krum~\cite{krum}, RFA~\cite{rfa}, ARAGG~\cite{karimireddy2022byzantinerobust}, Foolsgold~\cite{foolsgold}, FLAME~\cite{flame}, RLR~\cite{rlr}, and FLDectector~\cite{zhang2022fldetector}. 

\noindent\textbf{Evaluation metrics. }
We evaluate the model performance using the Mean Absolute Error (MAE) and Mean Squared Error (MSE). 
The considered metrics are defined as follows: 
$\text{MSE} = \frac{1}{n} \sum_{i=1}^{n} (y_i - \hat{y}_i)^2$; and
$\text{MAE} = \frac{1}{n} \sum_{i=1}^{n} |y_i - \hat{y}_i|$. The objective of the defender is to minimize both MSE and MAE.

\noindent\textbf{Simulation setting. }
We simulate an FL system with a total of \( N \) clients; in each training round, the server randomly selects \( k\% \) of clients to participate, a process known as client sampling. Unless otherwise specified, \( N \) is set to 30 and 100 for the PDCCH and FHWA datasets, respectively, with \( k \) fixed at 50\%. The number of communication rounds is set to 20 for both datasets, with 3 local training epochs per round. For all experiments and algorithms, we use SGD with consistent learning rates and a batch size of 128. A vanilla RNN model is employed as the backbone network. 
\subsection{Experimental Results}
\begin{figure*}[t!]
    \centering
    \begin{minipage}{0.46\textwidth}
        \captionof{table}{Defense efficacy under Byzantine Attacks with FHWA and PDCCH datasets.}
\scriptsize
\label{tab:byzantine_full}
\resizebox{0.95\linewidth}{!}{
\begin{tabular}{@{}l|cc|cc||cc|cc@{}}
\toprule
                         & \multicolumn{4}{c}{PDCCH Dataset}                         & \multicolumn{4}{c}{FHWA Dataset}                       \\ \cmidrule(lr){2-5} \cmidrule(lr){6-9}
      Methods                   & \multicolumn{2}{c}{Full} & \multicolumn{2}{c}{Partial} & \multicolumn{2}{c}{Full} & \multicolumn{2}{c}{Partial} 
                         \\
                         \cmidrule(lr){2-3} \cmidrule(lr){4-5} \cmidrule(lr){6-7} \cmidrule(lr){8-9}
                         
                         & MSE         & MAE        & MSE          & MAE          & MSE         & MAE        & MSE           & MAE         \\ \cmidrule(){1-9}
Krum                     & .0066       & .0455      & .0066        & .0478        & .0139       & .0741      & .0107         & .0661       \\
Multi-Krum               & .8706       & .7428      & .0105        & .0910        & .5991       & .6316      & 6.326         & 2.062       \\
RFA                      & .0053       & .0389      & .0071        & .0694     & \second{.0027}       
& \best{.0338}      & \second{.0052}         & \second{.0495}       \\
\update{ARAGG}                
& \update{.0056}         & \update{.0518}     
& \update{.0071}         & \update{.0576}  
& \update{.0252}         & \update{.1132}       
& \update{.0487}         & \update{.1691}       \\
FoolsGold                & .0591       & .1650      & .0575        & .2912        & .7358       & .7491      & .4878         & .6096       \\
FLAME                    & \second{.0047}       & \second{.0387}      & \second{.0058}        & \second{.0563}        & .0134       & .0703      & .0133         & .0712       \\
RLR                      & 33.21       & 5.545      & .0111        & .0652        & .3283       & .4439      & 1.488         & .9757       \\
FLDetector               & .0168       & .0944      & .0078        & .0750        & .3959       & .4986      & .0295         & .1159       \\ \cmidrule(){1-9}
\rowcolor{Gray}
\cellcolor{white}
Ours                     
&
\best{.0045}       
& \best{.0378}      
& \best{.0045}        
& \best{.0376}        
& \best{.0021}       
& \second{.0351}      
& \best{.0011}         
& \best{.0227}       \\ \bottomrule
\end{tabular}
}

    \end{minipage}
    \hfill
    \begin{minipage}{0.52\textwidth}
    \includegraphics[width=0.95\textwidth]{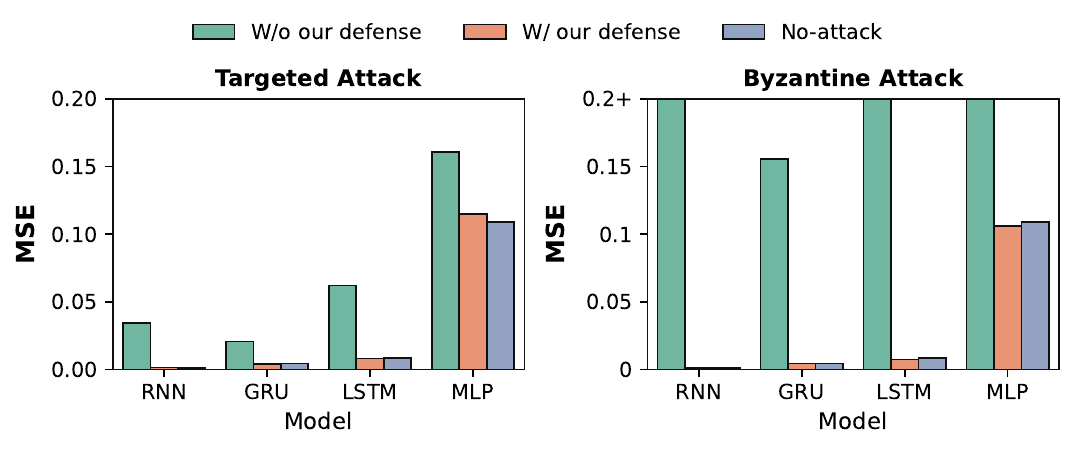}
    \vspace{-0.15cm}
    
    \caption{Zoom-in robustness of our method when applied with different model architectures.}
    \label{fig:change-model}
    \end{minipage}
\end{figure*}

\begin{table*}[tb!]
\centering
\caption{Defense efficacy under targeted attacks with different model poisoning techniques under the PDCCH and FHWA datasets.}
\label{tab:targeted_5G}
\resizebox{0.99\textwidth}{!}{

\begin{tabular}{@{}l|cc|cc|cc|cc||cc|cc|cc|cc@{}}
\toprule
\multirow{3}{*}{Methods} & \multicolumn{8}{c}{PDCCH}                                                                                 & \multicolumn{8}{c}{FHWA}                                                                                  \\ \cmidrule(lr){2-9} \cmidrule(lr){10-17}
                         & \multicolumn{2}{c}{Targeted} & \multicolumn{2}{c}{PGD} & \multicolumn{2}{c}{CnS} & \multicolumn{2}{c}{MR} & \multicolumn{2}{c}{Targeted} & \multicolumn{2}{c}{PGD} & \multicolumn{2}{c}{CnS} & \multicolumn{2}{c}{MR} \\
                         \cmidrule(lr){2-3} \cmidrule(lr){4-5} \cmidrule(lr){6-7} \cmidrule(lr){8-9} \cmidrule(lr){10-11} \cmidrule(lr){12-13} \cmidrule(lr){14-15} \cmidrule(lr){16-17}
                         & MSE ($\downarrow$)          & MAE ($\downarrow$)          & MSE ($\downarrow$)        & MAE ($\downarrow$)        & MSE  ($\downarrow$)        & MAE ($\downarrow$)        & MSE ($\downarrow$)        & MAE ($\downarrow$)       & MSE ($\downarrow$)           & MAE ($\downarrow$)          & MSE ($\downarrow$)        & MAE ($\downarrow$)        & MSE ($\downarrow$)        & MAE ($\downarrow$)        & MSE ($\downarrow$)        & MAE ($\downarrow$)       \\ \midrule
Krum                     
& .0066         & .0478        & .0066      & .0478      & .0107      & .0854      & .0066      & .0472
& .0257    & .1021     
& .0257    & .1021   
& .0049    & .0475       
& .0226    & .0882      \\
Multi-Krum               & .0105         & .0910        & .0105      & .0910      & .0058      & .0550      & .0105      & .0909     
& .0279    & .1377    
& .0279    & .1377  
& 1.001    & .9080        
& .0287    & .1396      \\
RFA                      
& \second{.0071}     & \second{.0694}    
& \second{.0071}     & \second{.0694}  
& .0059      & .0581      & .0072      & .0695     
& \second{.0126}       & \second{.0823}    
& \second{.0135}       & \second{.0874}  
& \second{.0135}       & \second{.0874}       
& \second{.0124}       & \second{.0805}      \\
ARAGG                    & .0106         & .0669        & .0076      & .0520      & .0065      & .0491      & .0127      & .0856     
& .2211        & .4215    
& .2433        & .4366        
& .5772        & .6744	       
& .2207        & .4212      \\
FoolsGold                & .0575         & .2091        & .0575      & .2091      & .0521      & .2155      & .0568      & .0564     
& .1333         & .5459    
& .1333        & .2960  
& 2.394         & 1.368       
& .1003         & .2576      \\
FLAME                    & .0058         & .0563        & .0058      & .0563      & .0053      & .0482      & \second{.0058}   & .0563      
& .0297         & .1221    
& .0297         & .1221  
& .0358         & .1397        
& .0297         & .1221      \\
RLR                      & .0111         & .0652        & .0111      & .0652      & .0069      & \second{.0466}      & .0061      & .0456     
& .0254         & .1154    
& .0254         & .1154  
& 1.325         & 1.032       
& .0260         & .1214      \\
FLDetector               
& .0078         & .0750        
& .0078      & .0750      
& \second{.0048}    & \second{.0460}        
& .0080      & .0783     
& .0233       & .1239    
& .0233       & .1239  
& .0194       & .0971       
& .0395       & .1647      \\ \midrule

\rowcolor{Gray}
\cellcolor{white}
Ours                     
& \best{.0046}  & \best{.0380}    
& \best{.0046}  & \best{.0380}  
& \best{.0046}  & \best{.0385}       
& \best{.0046}  & \best{.0380}      
& \best{.0014}       & \best{.0318}    
& \best{.0014}       & \best{.0318}  
& \best{.0012}       & \best{.0261}       
& \best{.0014}       & \best{.0318} \\ \bottomrule
\end{tabular}
}
\end{table*}

\subsubsection{\textbf{RQ1:} Can \thename{} robustly defend against both untargeted and targeted poisoning attacks? }First, we report the effectiveness of our defense and other baselines against both untargeted and targeted attacks, measured by MSE and MAE metrics, in Table~\ref{tab:byzantine_full} and Table~\ref{tab:targeted_5G}. \\
\smbf{Robustness on Untargeted Attack.}
\label{subsec:untargeted}
We first evaluate our method against Byzantine attacks and compare it with state-of-the-art defenses under full and partial participation scenarios. In the full scenario, all clients participate in training, while in the partial scenario, the server $\mathcal{S}$ randomly selects $m$ clients each round.
Results are shown in Table~\ref{tab:byzantine_full}, with the best highlighted in bold. Most defenses fail to mitigate the effect of malicious clients. Our method achieves the best performance in both settings across the PDCCH and FHWA datasets, reducing MSE by up to 78.84\% compared to the second-best baseline.
In contrast, other baselines, except RFA, show notable degradation on the FHWA dataset, a more practical and heterogeneous scenario. Since FTS differs from classification tasks, client data may exhibit diverse logical patterns, and even benign clients’ gradients are not necessarily close.
Moreover, methods such as Krum, RFA, FLAME, and ARAGG perform reasonably well in the full setting but degrade in partial scenarios. Overall, our method demonstrates superior generalization, maintaining robust performance while others show variability or degradation, especially when only a subset of clients participates each round.

\begin{figure}[t]
    \centering
        \centering
        \includegraphics[width=1.0\linewidth]{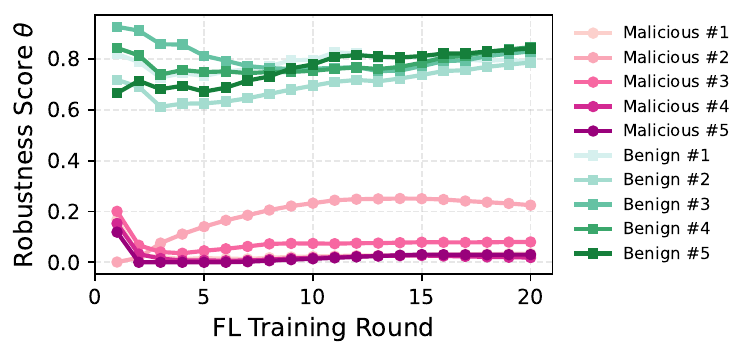}
    \caption{Selected clients' robustness scores $\theta$ over training rounds on the FHWA dataset.}
    \label{fig:selected-scores-by-round}
    \vspace{-0.6cm}
\end{figure}

\smbf{Robustness on Targeted Attack.}
\label{subsec:targeted}
Table~\ref{tab:targeted_5G} demonstrates our method's effectiveness under four white-box targeted attacks: Targeted, PGD, Constrain-and-scale (CnS), and Model Replacement. Our method consistently achieves the lowest MSE and MAE across all settings, outperforming the second-best baseline by 20.69\% on PDCCH and 93.27\% on FHWA in MSE. Existing baselines exhibit unstable performance across datasets and degrade under higher heterogeneity, as pairwise distance-based detection becomes ineffective when even benign updates deviate significantly. When adversaries combine data and model poisoning, baselines show mixed results: RLR mitigates CnS but struggles with Targeted and PGD attacks, while FLDetector handles CnS better than PGD or Model Replacement. Defenses relying on fixed metrics such as Euclidean or cosine distance remain susceptible to white-box attacks. In contrast, our method maintains robust performance across all scenarios. 
As shown in~\autoref{fig:selected-scores-by-round}, benign clients consistently achieve higher and more stable robustness scores, while malicious clients exhibit lower, fluctuating scores, a divergence that grows over training rounds, confirming the score's effectiveness in detecting malicious clients.

\noindent\textbf{RQ1 Summary.} \thename{} shows high effectiveness, consistently surpassing existing defenses against both untargeted and targeted attacks, even when adversaries combine multiple model poisoning strategies to strengthen their impact.

\subsubsection{\textbf{RQ2:} Does \thename{} generalize across diverse model architectures and FL aggregators? }
\begin{figure*}[tb!]
    \centering
    \begin{subfigure}[t]{1.0\linewidth}
    \centering
     \includegraphics[width=0.25\linewidth]{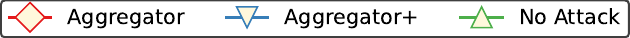}
    \end{subfigure}
    \begin{subfigure}[t]{0.18\linewidth}
    \includegraphics[width=1.0\linewidth]{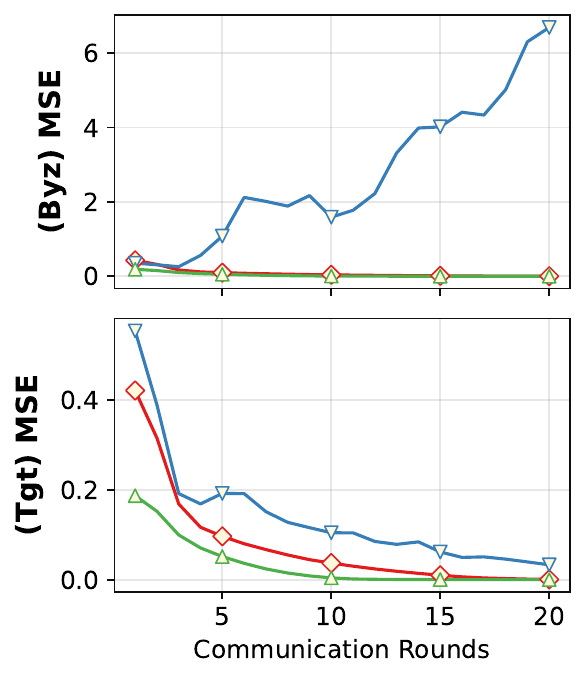}
    \vspace{-0.1cm}
    \caption{\scriptsize FedAvg}
    \end{subfigure}
    \begin{subfigure}[t]{0.18\linewidth}
    \includegraphics[width=1.0\linewidth]{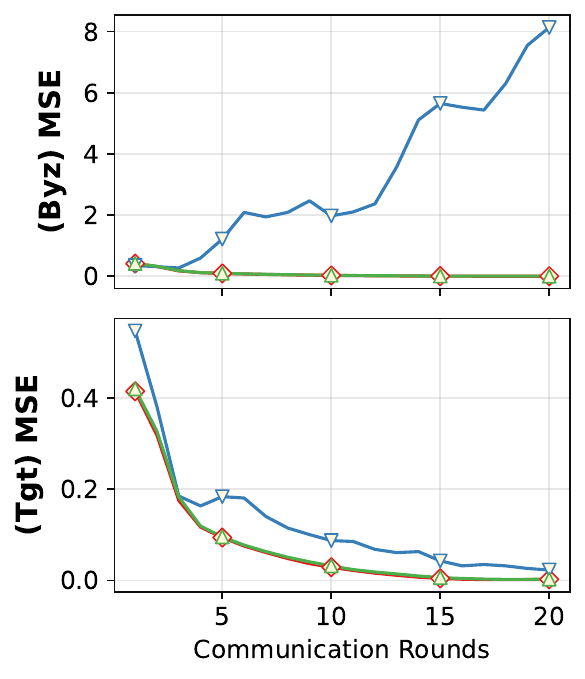}
    \vspace{-0.1cm}
    \caption{\scriptsize FedProx}
    \end{subfigure}
    \begin{subfigure}[t]{0.18\linewidth}
    \includegraphics[width=1.0\linewidth]{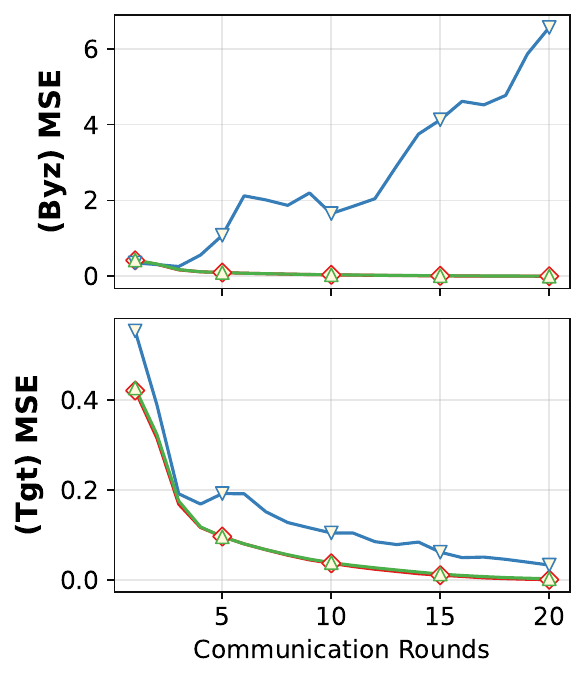}
    \vspace{-0.1cm}
    \caption{\scriptsize FedNova}
    \end{subfigure}
    \begin{subfigure}[t]{0.18\linewidth}
    \includegraphics[width=1.0\linewidth]{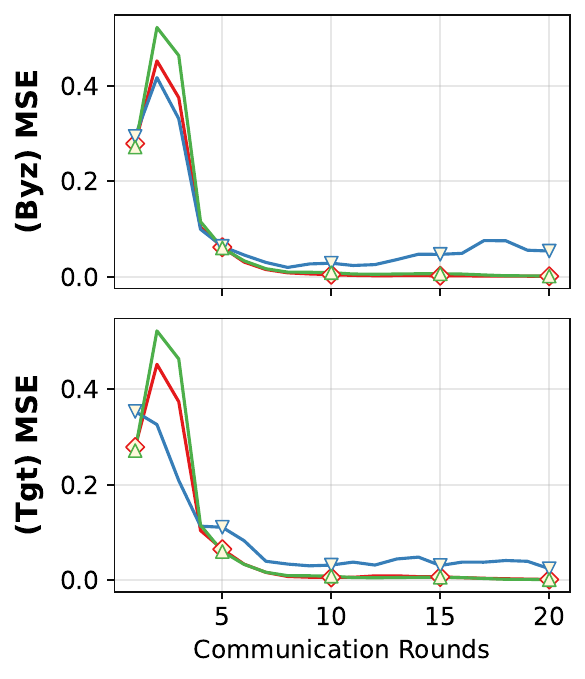}
    \vspace{-0.1cm}
    \caption{\scriptsize FedDyn}
    \end{subfigure}
    \begin{subfigure}[t]{0.18\linewidth}
    \includegraphics[width=1.0\linewidth]{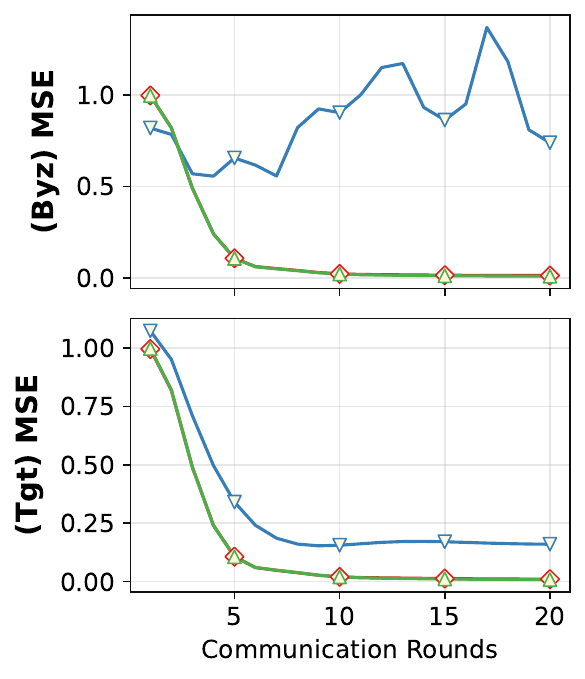}
    \vspace{-0.1cm}
    \caption{\scriptsize Scaffold}
    \end{subfigure}
    \vspace{-0.1cm}
    \caption{Evaluation of defense efficacy when integrating our method with various aggregators. \texttt{Aggregator$+$} indicates the combination of our method with the corresponding aggregator.}
    \label{fig:aggregators}
\end{figure*}

To answer this question, we evaluate the stability of our method across different FL configurations by varying two key factors that have been shown to strongly influence defense performance in FL~\cite{nguyen2024iba,flame}: (i) the model architecture and (ii) the FL aggregator, as illustrated in Figure~\ref{fig:change-model} and Figure~\ref{fig:aggregators}.

\noindent\textbf{Effect of Different Model Architectures.}
In this experiment, we assess different model architectures to verify the applicability of our method across various settings. We evaluated three versions for each architecture: the vanilla model (no attack), the poisoned model (attack without defense), and the defended model (attack with defense \thename{} applied) and presented the results in Figure~\ref{fig:change-model}.
As shown, adding \thename{} substantially mitigates poisoning effects caused by malicious clients, thus reducing MSE and MAE by over 99.6\% and 93.3\%, respectively. 
In the targeted attack scenario, \thename{} also shows its efficacy with different model architectures. 
Indeed, while the poisoned model exhibits performance degradation, using \thename{} with an LSTM backbone reduces the MSE by 66.7\% and the MAE by 59.1\%. 
In all model architectures, \thename{} consistently mitigates the effect of poisoning attacks as close to the one of a vanilla model. 
To this end, \thename{} can mitigate both untargeted and targeted attacks and provide robust performance improvements across various architectures.
\begin{table*}[tbh!]
\scriptsize
\captionof{table}{Comparing the performance of different defenses under various attack ratios $\epsilon$ with PDCCH and FHWA datasets.}
\label{tab:change_ratio}
\resizebox{1.0\linewidth}{!}{

\begin{tabular}{@{}l|cc|cc|cc|cc|cc||cc|cc|cc|cc|cc@{}}
\toprule
\multirow{3}{*}{Methods} & \multicolumn{10}{c}{PDCCH}                                                                                                       & \multicolumn{10}{c}{FHWA}                                                                                                        \\ \cmidrule(lr){2-11} \cmidrule(l){12-21} 
                         & \multicolumn{2}{c}{$\epsilon = $ 0.05} & \multicolumn{2}{c}{$\epsilon = $ 0.1} & \multicolumn{2}{c}{$\epsilon = $ 0.2} & \multicolumn{2}{c}{$\epsilon = $ 0.3} & \multicolumn{2}{c}{$\epsilon = $ 0.5} & \multicolumn{2}{c}{$\epsilon = $ 0.05} & \multicolumn{2}{c}{$\epsilon = $ 0.1} & \multicolumn{2}{c}{$\epsilon = $ 0.2} & \multicolumn{2}{c}{$\epsilon = $ 0.3} & \multicolumn{2}{c}{$\epsilon = $ 0.5} \\ 
                         \cmidrule(lr){2-3} \cmidrule(lr){4-5}
                         \cmidrule(lr){6-7} \cmidrule(lr){8-9}
                         \cmidrule(lr){10-11} \cmidrule(lr){12-13}
                         \cmidrule(lr){14-15} \cmidrule(lr){16-17}
                         \cmidrule(lr){18-19} \cmidrule(lr){20-21}

                         & MSE         & MAE        & MSE        & MAE        & MSE        & MAE        & MSE        & MAE        & MSE        & MAE        & MSE         & MAE        & MSE        & MAE        & MSE        & MAE        & MSE        & MAE        & MSE        & MAE        \\ \cmidrule(r){1-21}
Krum                     
& .0078       & .0505      
& .0070      & .0485      
& .0051      & .0435      
& .0072      & .0496      
& .0068      & .0425      
& \multicolumn{1}{r}{.0268} & .0965                   & \multicolumn{1}{r}{.0090} & .0619                   & \multicolumn{1}{r}{.0170} & .0661                   & .0031                   & .0356                   & \multicolumn{1}{r}{\second{.0065}} & \second{.0526}      \\
Multi-Krum               
& \best{.0044}       & \best{.0379}      & .2987      & .3869      & .7665      & .7005      & 1.178      & .8686      & 7.643      & 2.181          
& .1347       & .2897      
& 1.277      & .8875      & 6.326      & .0622      & 7.730      & 2.227      & 18.05      & 3.321      \\
RFA                      
 & .0048       & .0388      & .0051      & .0392      & .0053      & .0398      & .0052      & .0395      & \second{.0048}      & \best{.0388}     
& \multicolumn{1}{r}{.0045} & .0458                   & \multicolumn{1}{r}{\second{.0036}} & \second{.0382}                   & \multicolumn{1}{r}{\second{.0052}} & \second{.0495}                   & \second{.0037}                   & \second{.0426}                   & \multicolumn{1}{r}{.0069} & .0547      \\
ARAGG                    & .0063       & .0480      & .0067      & .0536      & .0106      & .0669      & .0138      & 1.037      & .0138      & .0902      & .0067       & .0691      & .0375      & .1574      & .2211      & .4215      & .3451      & .5220      & .8945      & .8493      \\
FoolsGold                & .0076       & .0492      & .0080      & .0605      & .0765      & .1950      & .0551      & .1594      & .0424      & .1394      & .0746       & .2304      & .3137      & .4797      & .4878      & .6096      & .6487      & .7124      & .9113      & .8498      \\
FLAME                    & \second{.0046}       & .0394      & .0047      & .0389      & \second{.0050}      & \second{.0391}      & \second{.0050}      & \second{.0393}      & .0780      & .2025     
& .0095       & .0626      & .0087      & .0580      & .0133      & .0712      & .0153      & .0753      & 5.844      & 1.969      \\
RLR                      & .0978       & .2679      & 2.011      & 1.117      & 2.515      & 1.403      & 2.456      & 1.367      & 39.88      & 5.820      & .1820       & .3551      & 1.328      & .9699      & 1.488      & .9757      & 7.001      & 2.096      & 3.689      & 1.562      \\
FLDetector               & .0047       & .0393      & \best{.0043}      & \best{.0408}      & 1.862      & 1.170      & 1.788      & 1.156      & 10.97      & 2.650

& \multicolumn{1}{r}{\second{.0026}} & \second{.0347}                   & \multicolumn{1}{r}{.0930} & .2117                   & \multicolumn{1}{r}{.0295} & .1159                   & .0612                   & .1813                   & \multicolumn{1}{r}{9.460} & 2.426      \\
\cmidrule(lr){1-21}
\rowcolor{Gray}
\cellcolor{white}
Ours                      
 & .0047       & \second{.0384}      
 & \second{.0047}      & \second{.0387}      
 & \best{.0045}      & \best{.0376}      
 & \best{.0045}      & \best{.0382}      
 & \best{.0047}      & \second{.0395}     
& \multicolumn{1}{r}{\best{.0014}} & \best{.0267}                   
& \multicolumn{1}{r}{\best{.0017}} & \best{.0285}                   
& \multicolumn{1}{r}{\best{.0011}} & \best{.0227}                   
& \best{.0011}                     & \best{.0243}                   
& \multicolumn{1}{r}{\best{.0019}} & \best{.0324}       \\ \bottomrule
\end{tabular}
}
\end{table*}

\begin{table}[tbh]
\centering
\caption{\update{\thename{} on adaptive attacks. The results are obtained with the default setting we use in the main experiment.}}
\label{tab:adaptive-attack}
\resizebox{0.8\linewidth}{!}{
\begin{tabular}{@{}c|cc|cc@{}}
\toprule
\multicolumn{1}{l}{\multirow{2}{*}{Ratio}} & \multicolumn{2}{c}{MSE} & \multicolumn{2}{c}{MAE} \\ \cmidrule(lr){2-3} \cmidrule(lr){4-5} 
\multicolumn{1}{l}{}                       & Adaptive  & No-Adaptive & Adaptive  & No-Adaptive \\ \cmidrule{1-5} 
0.05                                       & 0.0017    & 0.0014      & 0.0291    & 0.0267      \\
0.1                                        & 0.0015    & 0.0017      & 0.0260    & 0.0285      \\
0.2                                        & 0.0014    & 0.0011      & 0.0274    & 0.0227      \\
0.3                                        & 0.0023    & 0.0011      & 0.0332    & 0.0243      \\
0.5                                        & 0.0026    & 0.0019      & 0.0362    & 0.0324      \\ \bottomrule
\end{tabular}
}
\vspace{-0.5cm}
\end{table}

\noindent\textbf{Effect of Different FL Aggregators.}
In this experiment, we evaluate the performance \thename{} when combined with various FL aggregators with Byzantine and targeted attacks. 
We select five aggregators, including FedAvg, FedProx~\cite{fedprox}, Scaffold~\cite{karimireddy2020scaffold}, FedDyn~\cite{acar2021federated}, FedNova~\cite{wang2020tackling}, and denote their \thename-enhanced versions as Aggregator+.
The result is presented in Figure~\ref{fig:aggregators}, which reveals a sustainable contrast in performance between baseline methods and their corresponding \thename{}-enhanced versions. 
When integrated with different baselines, \thename{} consistently mitigated errors from attacks. Across all aggregators, it reduced MSE by over 99\%, yielding error levels close to the no-attack case.

\noindent\textbf{RQ2 Summary.}~\thename{} can be seamlessly integrated with different architectures and aggregation strategies while maintaining stable and robust defense performance.

\subsubsection{\textbf{RQ3:} How does \thename{} behave under varying adversarial strengths, including different attack ratios and strategies?}
We assess the robustness of~\thename{} under different adversarial strengths, where the attack ratio denotes the fraction of compromised clients in the federation; a higher ratio corresponds to a stronger adversary. In addition, we study a challenging \textbf{adaptive attack} in which the adversary is assumed to know the server-side defense strategy and crafts its poisoned updates accordingly to maximize the chance of bypassing our logic-guided verification.

\smbf{Effect of Different Attack Ratios. }
\label{subsec:ratio}
 Table~\ref{tab:change_ratio} demonstrates the efficacy of different defenses against various backdoor attacks under varying attack ratios $\epsilon \in [0.05, 0.1, 0.2, 0.3, 0.5]$ on both PDCCH and FHWA datasets. As $\epsilon$ increases, the scenario becomes more challenging due to the decreasing ratio of benign to malicious clients, with $\epsilon = 0.5$ being the most difficult case where benign and malicious clients are equal in number. The results reveal remarkable variations in baseline robustness: most methods deteriorate at higher ratios, with methods such as Multi-Krum and FoolsGold experiencing severe degradation, and Multi-Krum's MSE escalating more than 10 times as $\epsilon$ rises from 0.05 to 0.5. In contrast, \thename{} consistently achieves the lowest MSE and MAE across all attack ratios, maintaining stable performance even at $\epsilon = 0.5$, demonstrating its robustness and reliability.

\smbf{Adaptive Attacks. }
We evaluate~\thename{} under an adaptive attack setting, where we assume the adversary knows the defense mechanism employed by the server, enabling it to attempt to circumvent it. Since~\thename{}'s core defense lies in its ability to leverage global reasoning properties to evaluate clients' trustworthiness, we developed adaptive attacks based on the premise that adversaries could infer logical properties from the global model to obscure their suspicious behavior and constrain their models to produce predictions that satisfy these properties.
As shown in Table~\ref{tab:adaptive-attack}, adaptive attacks are largely ineffective against~\thename{}. Even under the most adversarially favorable scenario — 50\% of the clients are malicious — the attacks achieve only limited success in evading the defense, with errors still half those of the second-best baseline (RFA), a scenario generally considered impractical in typical federated learning systems. The failure of these adaptive attacks stems from a trade-off: as poisoned models attempt to satisfy the logical properties required for trustworthiness, they inadvertently cause malicious clients' models to align with the behavior of benign clients, diminishing the attack's impact on the global model. Thus, while designing an adaptive attack capable of evading~\thename{} is theoretically possible, it simultaneously introduces errors that undermine the adversary's objectives.

\noindent\textbf{RQ3 Summary.} \thename{} remains robust under varying adversarial strengths and adaptive settings, consistently achieving the lowest error across all attack ratios, even with 50\% malicious clients. Adaptive attacks that try to bypass the server defense also fail to compromise the defense.

\section{Conclusion}

In this study, we address a critical gap in FL robustness by focusing on the specific challenges posed by poisoning attacks within the FTS domain.
Mainstream FL defenses have primarily targeted computer vision applications and
showcase their limited efficacy with FTS.
We introduce~\thename{}, a defense mechanism designed to tackle these challenges effectively.
\thename{} leverages a unique approach that combines formal logic reasoning with hierarchical clustering to evaluate client trustworthiness and align predictions with global time-series patterns.
This method diverges from traditional model-centric defenses by focusing on the logical consistency of client contributions, which enhances its ability to detect and mitigate adversarial behaviors.
The results from our experiments on two distinct datasets and various attack settings demonstrate that~\thename{} 
outperforms existing baseline methods, confirming its efficacy in improving the robustness of FL systems for time series applications.

\section*{Acknowledgments}
This work was supported in part by the National Science Foundation under Grant Nos. 2443803 and 2427711, the NSA Science of Security (SoS) program, and the ARPA-H DIGIHEALS program.
Any opinions, findings, conclusions, or recommendations expressed in this material are those of the authors and do not necessarily reflect the views of the National Science Foundation or the U.S. Government.

\bibliographystyle{IEEEtran}
\bibliography{main}

\newpage
\appendix
This document serves as an extended exploration of our research, providing a detailed implementation and discussion of experiments provided in the main text. 
Appendix~\ref{appendix:stl} presents additional details on STL and the STL-based property inference and verification tasks.
Appendix~\ref{appendix:training_detail} provides a comprehensive discussion of the training process, including datasets, model structures, and configurations used to reproduce the reported results. Additionally, we present supplementary results not included in the main paper in Appendix~\ref{appendix:discussion}. 
\begin{table*}[tb!]

\centering
\caption{Examples of temporal reasoning templates specified with STL.}
\label{tab:stl-case}
\resizebox{0.95\linewidth}{!}{
\begin{tabular}{@{}llll@{}}
\toprule
\multicolumn{1}{l}{\textbf{Property}}                     
& \multicolumn{1}{l}{\textbf{Description}}                                                                                          & \multicolumn{1}{l}{\textbf{Templatized Logic Formula}}             
& \textbf{Parameters}   \\ \midrule
\multicolumn{1}{l}{\makecell[l]{Operational \\ Range}}          & \begin{tabular}[c]{@{}l@{}}Signal is upper-bounded by threshold $a$ \\ and lower-bounded by threshold $b$.\end{tabular}                                        & $\bigwedge_{i=1}^{1, 2, ..., \tau} (\square_{[i,i+t]} (x\leq a_i\wedge x\geq b_i) ) $      & $a_i, b_i $           \\
\cmidrule{1-4}
\multicolumn{1}{l}{Existence}                                                              & \begin{tabular}[c]{@{}l@{}}Signal should eventually reach \\ the upper extreme $a$ and the lower extreme $b$.\end{tabular}                                     & $\bigwedge_{i=1}^{1, 2, ..., \tau} (\Diamond_{[i,i+t]} (x \leq a_i \wedge x \geq b_i) ) $  & $a_i, b_i$            \\ \cmidrule{1-4}
\multicolumn{1}{l}{Until}                                                                  & \begin{tabular}[c]{@{}l@{}}Signal must satisfy one specification \\ at all times until another condition is met.\end{tabular}                                  & $\bigwedge_{i=1}^{1, 2, ..., t} ((x < a_i) \, \mathcal{U}_{[i,i+1]} (x < b_i))$            & $a_i, b_i$            \\ \cmidrule{1-4}
\multicolumn{1}{l}{\makecell[l]{Intra-task \\ Reasoning}}        & \begin{tabular}[c]{@{}l@{}}The difference between signal variables $x_1$ and $x_2$ \\ should be greater than $a$.\end{tabular}                                 & $\bigwedge_{i=1}^{1, 2, ..., \tau} (\square_{[i,i+t]}((x_1-x_2)>a_i))$                     & $a_i$                 \\ \cmidrule{1-4}
\multicolumn{1}{l}{\makecell[l]{Temporal \\ Implications}}       & \begin{tabular}[c]{@{}l@{}}The happening of one event indicates that \\ another event will happen at some point in the future.\end{tabular}                    & $\square_{[t_1,t_2]} ((x \geq a_1) \rightarrow \Diamond_{[t_3,t_4]}(x \geq a_2))$          & $a_1, a_2$            \\ \cmidrule{1-4}
\multicolumn{1}{l}{\makecell[l]{Intra-task \\ Nested Reasoning}} & \begin{tabular}[c]{@{}l@{}}The signal variable $x_1$, when greater than a threshold $a$, \\ indicates $x_2$ will eventually reach a threshold $b$\end{tabular} & $\square_{[t_1,t_2]}((x_1 \geq a) \rightarrow \Diamond_{[t_3,t_4]} (x_2 \geq b))$          & $a, b$                \\ \cmidrule{1-4}
\multicolumn{1}{l}{\makecell[l]{Multiple \\ Eventualities}}      & Multiple events must eventually happen, but their order can be arbitrary.                                                                                      & $\Diamond_{[t_1,t_2]} (x \geq a_1) \wedge \cdots \wedge \Diamond_{[t_3,t_4]} (x \geq a_n)$ & $a_1, a_2 \cdots a_n$ \\ \cmidrule{1-4}
\multicolumn{1}{l}{\makecell[l]{Template-free}}                                                          & Specification mining without a templatized formula.                                                                                                              & No pre-defined templates are needed.                                                       & n/a.                  \\ \bottomrule
\end{tabular}
}
\end{table*}

\subsection{Signal Temporal Logic: Inference and Verification}
\label{appendix:stl}
To begin, we follow~\cite{an2024formal,maler2004monitoring} to present the qualitative (Boolean) semantics for an STL formula $\varphi$. Here, we use the following notations: $\mathbf{x}$ denotes a signal trace, $\mu$ denotes an STL predicate, and $\varphi, \varphi_1$, and $\varphi_2$ represent different STL formulas.
\update{In our implementation, we leverage open-source TeLEx~\cite{jha2017telex} to conduct logical specifications for time-series. The usage of this open-source framework is as follows:
\begin{itemize}
    \item Users define the templates containing the specification types for logical reasoning given a time series. The template should follow the grammar and syntax of STL. Example~\ref{eg:tempstl} is a good reference.
    \item Given the templates and a time series, TeLEx can output the corresponding parameters automatically using optimization methods. 
\end{itemize}
This framework only requires the knowledge of formal logic representation and formulas to represent desired formal formulas. Examples~\ref{eg:tempstl} and \ref{eg:verification} provide a detailed explanation and guidelines for describing an STL property. TeLEx finds the value of the unknown parameters such that the synthesized STL property is satisfied by all the provided traces and it is tight. Our method can be adapted to other STL mining methods or frameworks since the working mechanism will not change upon the framework changes.}
\begin{definition}[STL Qualitative Semantics]\label{def:qualitative} 
$$
\begin{array}{lll}
(\mathbf{x}, t) \vDash \top & \leftrightarrow & \top \\
(\mathbf{x}, t) \vDash \mu & \leftrightarrow & \mu(\mathbf{x}[t]) \\
(\mathbf{x}, t) \vDash \varphi_1 \vee \varphi_2 & \leftrightarrow & (\mathbf{x}, t) \vDash \varphi_1 \vee(\mathbf{x}, t) \vDash \varphi_2 \\
(\mathbf{x}, t) \vDash \varphi_1 \wedge \varphi_2 & \leftrightarrow & (\mathbf{x}, t) \vDash \varphi_1 \wedge(\mathbf{x}, t) \vDash \varphi_2 \\
(\mathbf{x}, t) \vDash \Diamond_{[a, b]} \varphi & \leftrightarrow & \exists t^{\prime} \in[t+a, t+b],\left(\mathbf{x}, t^{\prime}\right) \vDash \varphi \\
(\mathbf{x}, t) \vDash \square_{[a, b]} \varphi & \leftrightarrow & \forall t^{\prime} \in[t+a, t+b],\left(\mathbf{x}, t^{\prime}\right) \vDash \varphi \\
(\mathbf{x}, t) \vDash \varphi_1 \mathcal{U}_{[a, b]} \varphi_2 & \leftrightarrow & \exists t^{\prime} \in[t+a, t+b],\left(\mathbf{x}, t^{\prime}\right) \vDash \varphi_2 \\
& & \wedge \forall t^{\prime \prime} \in\left[t, t^{\prime}\right],\left(\mathbf{x}, t^{\prime \prime}\right) \vDash \varphi_1
\end{array}
$$
\end{definition}

While the qualitative semantics in Def.~\ref{def:qualitative} provide a Boolean evaluation of the satisfaction of an STL property, there are inference tasks that rely on a real-valued measurement of property satisfaction, known as the STL robustness metric $(\rho)$. Def.~\ref{def:quantitative} below describes how this metric is calculated, which maps a given signal trace $\mathbf{x}$ and an STL formula $\varphi$ to a real number over a specified time interval $I$.

\begin{definition}[STL Quantitative Semantics]\label{def:quantitative} 
The robustness metric $\rho$ maps an STL formula $\varphi$, a signal trace $\mathbf{x}$, and a time $t$ to a real value such that:
\begin{align*}
& \rho(\mathbf{x}, \varphi, t) &&= g(\mathbf{x}(t))-\alpha \text { where } \mu(X) \text { is } g(X) \geq \alpha\\
&\rho(\neg \varphi, \mathbf{x}, t) &&= - \rho(\varphi, \mathbf{x}, t) \\
&\rho(\varphi_1 \vee \varphi_2, \mathbf{x}, t) &&=  \max\{\rho(\varphi_1, \mathbf{x}, t), \rho(\varphi_2, \mathbf{x}, t) \}\\
&\rho(\varphi_1 \land \varphi_2, \mathbf{x}, t) &&=  \min\{\rho(\varphi_1, \mathbf{x}, t), \rho(\varphi_2, \mathbf{x}, t) \}\\
& \rho(\Diamond_I \varphi , \mathbf{x}, t) && = \underset{t' \in (t, t+I)}{\max} \rho(\varphi, \mathbf{x}, t')\\
& \rho(\square_I \varphi, \mathbf{x}, t) && = \underset{t' \in (t, t+I)}{\min} \rho(\varphi, \mathbf{x}, t') \\
&\rho(\varphi_1 \mathcal{U} \varphi_2, \mathbf{x}, t) &&= \sup_{t'\in (t + I) \cap \mathbb{T}} (\min\{\rho(\varphi_2, \mathbf{x}, t'), \\
&&& \, \quad \inf_{t''\in[t,t']}(\rho(\varphi_1, \mathbf{x}, t'')) \})
\end{align*}
\end{definition}
\subsection{Training configurations}
\label{appendix:training_detail}
\subsubsection{Baselines}
\begin{itemize}
    \item \textit{Krum/Multi-Krum~\cite{krum}: }Krum and Multi-Krum algorithms are Byzantine-resilient aggregation techniques designed for federated learning to defend against malicious or faulty clients during model training. These methods work by selecting the update(s) from a client(s) closest to most other clients’ updates, minimizing the impact of outliers or adversarial updates.
    \item \textit{RFA~\cite{rfa}: }RFA replaces the weighted averaging mechanism by using the geometric median for aggregating model updates, protecting against data and model poisoning without revealing individual contributions.
    \item  \update{\textit{ARAGG~\cite{karimireddy2022byzantinerobust}: }}\update{ARAGG introduces a guaranteed convergence approach for addressing non-iid Byzantine attacks. It leverages bucketing of local clients and incorporates combined worker momentum to derive robust updates by effectively utilizing historical information.}
    \item \textit{FoolsGold~\cite{foolsgold}: }FoolsGold adjusts the learning rate for each client based on updated similarity and historical data. It uses cosine similarity to measure the angular distance between updates.
    \item \textit{FLAME~\cite{flame}: }FLAME is a backdoor defense that includes three components: DP-based noise injection to remove backdoor contributions, unsupervised model clustering to detect and eliminate poisoned updates, and weight clipping to limit the impact of malicious updates. 
    \item \textit{RLR~\cite{rlr}: }RLR proposes a lightweight defense against backdoor attacks in federated learning by adjusting the aggregation server's learning rate per dimension and per round, based on the majority sign of agents' updates.
    \item \textit{FLDetetor~\cite{zhang2022fldetector}: }FLDetector identifies and removes malicious clients in federated learning by monitoring the consistency of their model updates. It predicts updates using the Cauchy mean value theorem and L-BFGS, flagging clients as malicious if their actual updates deviate from predictions over multiple iterations.
\end{itemize}

 With the implementations of these baselines, most hyper-parameters are inherited with minor modifications. Several parameters are adjusted to make experiment settings more appropriate. Other hyperparameters without mention are set up as in the original works.

\noindent\textbf{Robust threshold $\theta$ of RLR \cite{rlr}.} The value of $\theta$ in the RLR method is specified to be any value between [$m\cdot \epsilon+1$, $m-m\cdot \epsilon$], where $m$ is the number of participants each round and $\epsilon$ is the proportion of malicious clients, according to the authors. Because there is limited to one malicious client each round, the value of $\theta$ is set to $1$ throughout the experiments with all datasets.

\noindent\textbf{Estimated number of Byzantine clients $F$ in Krum/Multi-Krum \cite{krum}.} Since the experiments are conducted under fixed-pool backdoor attacks with a maximum of one malicious client, so $F$ is set to be $\max{(\lfloor \epsilon \cdot m \rfloor, 1)}$.

\noindent\textbf{Estimated number of Byzantine clients in FLDetector \cite{zhang2022fldetector}.} We set the number of byzantine clients as $\lfloor \epsilon \cdot m \rfloor$, corresponding to the potential number of malicious clients appearing in each training round.

\noindent\textbf{\update{Bucketing size and momentum of ARAGG. }}
\update{In experiments with ARAGG, we set the bucketing size to 2, as suggested in the paper. In addition, the coefficient of momentum $\beta$ is set to 0.5, which is used to reduce the impact of gradient variance from local clients.}

\subsubsection{Training hyper-parameters}
We have a fixed number of 30/100 FL clients for PDCCH/FHWA in each testing scenario. During the communication rounds, 50\% of the clients are randomly selected. The batch size is 128, and the test batch size is 256. We set the maximum learning rate for the MLP, RNN, LSTM, and GRU models to be 0.001. 
To optimize the models, we utilize the SGD optimizer in the PyTorch implementation. 
We use the mean squared error (MSE) loss, implemented in PyTorch, which is commonly employed in regression tasks. In our default setup, the number of local training epochs is set to 3 if not otherwise specified.

\subsubsection{Different Aggregators}
\begin{itemize}
    \item \textit{FedProx~\cite{fedprox}: }An extension of FedAvg that incorporates a proximal term to ensure that local models remain close to the global model, enabling better performance in non-IID settings and allowing for partial client participation.
    \item \textit{FedDyn~\cite{acar2021federated}: }Introduces a dynamic regularization technique for each device in Federated Learning to ensure alignment between local and global solutions over time. This approach improves training efficiency in both convex and non-convex settings while being robust to device heterogeneity, partial participation, and unbalanced data.
    \item \textit{Scaffold~\cite{karimireddy2020scaffold}: }Aims to mitigate the drift between local and global models by using control variates, which stabilize client updates and enhance convergence, particularly in settings with high data heterogeneity
    \item \textit{FedNova~\cite{wang2020tackling}: }Enhances the federated averaging process by considering the contribution of each client based on the number of samples they have, aiming to improve convergence rates and model performance, especially when clients have varying amounts of data
\end{itemize}

\noindent\textbf{Hyper-parameters for aggregators. }With FedProx, we set the proximal term scaled $\mu$ to 0.01, followed~\cite{perifanis2023federated} which restricts the trajectory of the iterates by constraining the iterates to be closer to that of the global model. In FedDyn, the dynamic regularization coefficient $\alpha$ is set to 0.1, which dynamically modifies local loss functions so that local models converge to a consensus consistent with stationary points of the global loss. In FedNova, the local momentum factor $\rho$ is 0.1, and this parameter helps to control the SGD optimization during the local training process and reduce cross-client variance. Other parameters without further mention are inherited from original implementations.

\subsubsection{\update{Different Model Architectures}}
\update{
We summarized the model configuration in Table~\ref{tab:model-config}. 
This setting is applied for both datasets; given a specific time-series data, the input and output dimensions will be adapted correspondingly. Please refer to our public implementation for more implementation details.}
\begin{table}[tbh]
\centering
\caption{\update{Model configurations for different models used with \thename{}.}}
\label{tab:model-config}
\resizebox{1.0\linewidth}{!}{
\begin{tabular}{@{}llcccc@{}}
\toprule
\textbf{Model} & \textbf{Layers} & \textbf{Hidden Dims} & \textbf{Batch Size} & \textbf{Learning Rate} & \textbf{Optimizer} \\ \midrule
MLP            & 3 Hidden + 1 FC & {[}256, 128, 64{]}   & 128                 & 0.001                  & Adam               \\
RNN            & 1 RNN + MLP     & 128                  & 128                 & 0.001                  & Adam               \\
LSTM           & 1 LSTM + MLP    & 128                  & 128                 & 0.001                  & Adam               \\
GRU            & 1 GRU + MLP     & 128                  & 128                 & 0.001                  & Adam               \\ \bottomrule
\end{tabular}
}
\end{table}

\subsubsection{\thename{}'s hyper-parameters}
In~\thename{}'s method, the parameter $\gamma$ in Eqn.~\ref{eqn:update-mask} plays an important role in detecting malicious clients. Specifically, clients whose scores fall below $\gamma$ fraction of the highest robustness score across all clients are considered malicious and are filtered out. This technique is based on the premise that adversarial or poisoned models typically exhibit lower robustness scores than the majority of benign clients, which allows them to be detected and excluded from the aggregation process. This parameter should be set close to the number of malicious clients in each training round to balance the trade-off between detecting poisoning updates and maintaining performance on the main task in FL. In our experiments, we the $\gamma$ to 0.2 with the FHWA dataset and $\gamma$ to 0.5 with the PDCCH dataset. Following Krum and FLDectector, this parameter should be set to be in the range of $[\hat{\epsilon}, 0.5]$, --- the estimated fraction of poisoning clients.

\noindent\textbf{\update{Effect of $\gamma$ parameter on \thename{}. }}
\update{A discussion about the importance of the threshold $\gamma$ appears only in the appendix; it's mentioned that it should be set close to the number of malicious clients in each round, which might not be realistic to be known. Then it'd be interesting to show how the performance of FLORAL depends on such a threshold in practice.}
\begin{figure}[tbh]
    \centering
\includegraphics[width=0.85\linewidth]{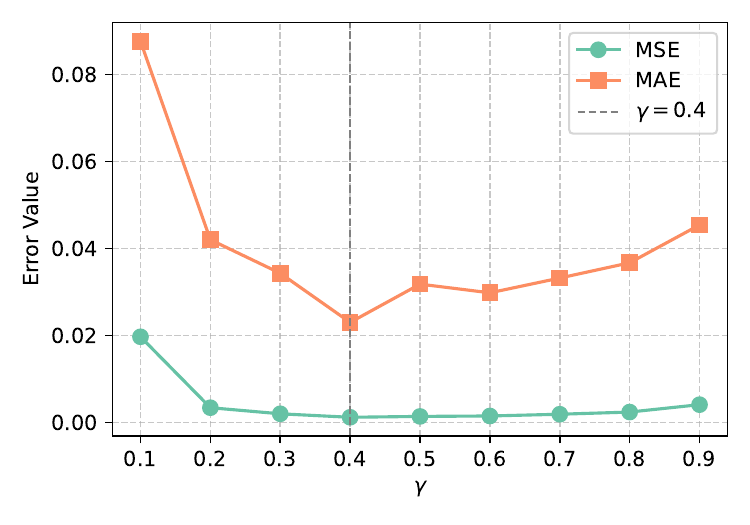}
    \caption{\update{Effect of parameter $\gamma$ on the performance of \thename{}.}}
    \label{fig:gamma-study}
\end{figure}

\subsection{\update{Inference Time Analysis}}
\update{
Table~\ref{tab:inference_verification} shows the average time required for extracting and verifying various types of logic reasoning properties across all clients in the FL system.~\footnote{We ignore the deviation in verification due to it is approximately zeros.} 
The provided runtimes represent the average time taken to extract each of the logic properties as outlined in Table~\ref{tab:stl-case} of the paper. 
Notably, for straightforward yet expressive properties like operational range and existence, the average extraction time for one property on a time series with a length of 2400 can be as short as 0.00320 seconds. Conversely, the most time-consuming property to extract is multiple eventualities, primarily due to the complexity of the multiple $\land$ operations involved.
By enumerating the number of property formulas per client, we generally observe a linear trend in the time increment. 
The average inference/mining time per property within our operational range is approximately 0.003 seconds, allowing the server to compute about 333.33 formulas per second; and the verification time is much faster compared to the inference time. This demonstrates the high efficiency and practicality of our approach.
It is worth noting that while the computation time increases linearly with the number of clients in each training round if the properties are processed sequentially, this cost can be substantially mitigated through parallelization, provided the server has sufficient computational resources. Modern server infrastructures are often designed to support such parallel processing, making this optimization feasible and further reducing the overhead.
Compared to the communication costs, which are the primary bottleneck in federated learning systems, the computational overhead introduced by our method is minimal. Additionally, many federated defense studies prioritize performance improvements without explicitly addressing runtime costs. The runtime requirements of our method, which typically amount to only a few seconds per round, are well within acceptable limits for FL scenarios.
In summary, our approach achieves a balance between computational efficiency and robust adversarial defense, demonstrating its scalability and compatibility with real-world FL systems without imposing significant additional complexity.}
\begin{table}[tbh!]
\centering
\caption{\update{Inference and verification times for different logical properties.}}
\resizebox{0.95\linewidth}{!}{
\begin{tabular}{@{}lcc@{}}
\toprule
\textbf{Properties}         & \textbf{Inference Time (s)} & \textbf{Verification Time (s)} \\ \midrule
Operational Range           & 0.0030 $\pm$ 0.0001         & 0.000252                       \\
Until                       & 0.1179 $\pm$ 0.0900         & 0.001188                       \\
Existence                   & 0.0188 $\pm$ 0.0023         & 0.000346                       \\
Intra-task Reasoning        & 0.0107 $\pm$ 0.0005         & 0.000336                       \\
Temporal Implications       & 0.0186 $\pm$ 0.0005        & 0.001223                       \\
Intra-task Nested Reasoning & 0.0195 $\pm$ 0.0007         & 0.001148                       \\
Multiple Eventualities      & 0.1807 $\pm$ 0.0101         & 0.002870                        \\ \bottomrule
\end{tabular}}
\label{tab:inference_verification}
\end{table}

\subsection{Discussion and Limitations}
\label{appendix:discussion}
\begin{figure}[hbt!]
    \centering
    \includegraphics[width=0.8\linewidth]{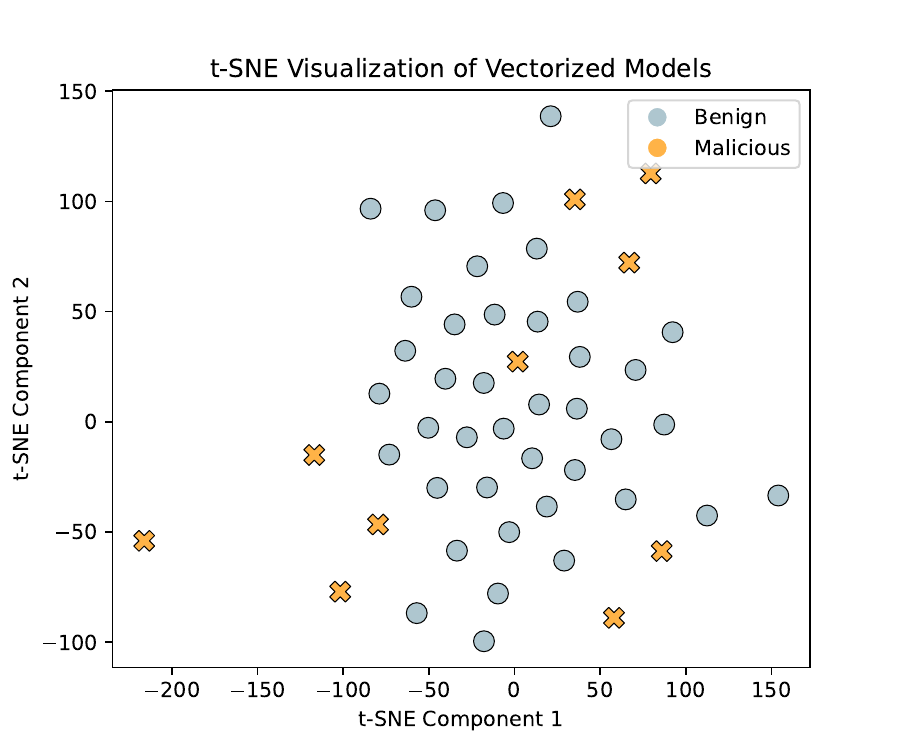}
    \caption{\update{Visualization of local updates of benign and malicious clients.}}
    \label{fig:enter-label}
\end{figure}
\update{The primary challenges posed by Federated Time Series (FTS) data to existing robustness methods stem from its inherent characteristics. Unlike image datasets for classification, where benign models typically converge due to shared features and objectives, time series data often involve unique patterns or logical rules specific to each client. This heterogeneity hinders alignment among benign models, especially in earlier training rounds, allowing poisoned models to emerge as close neighbors to benign ones. Furthermore, missing malicious clients disrupt the continuity of training, causing models to deviate from expected behaviors and propagating these deviations across subsequent rounds, ultimately impacting the training of benign clients. Existing methods like RLR (Reversing Layer Regularization), which depend on gradient sign agreement, struggle to adapt to the heterogeneity and logical variations in time series data, often failing to identify a global solution. Similarly, clustering-based approaches such as FLAME and FLDetector are ineffective at separating benign and poisoned models due to the high variance among benign models. These limitations highlight that model-centric solutions alone are insufficient for the time-series domain, motivating the exploration of logical properties, which have proven successful in capturing the intrinsic characteristics of time series data.}

To address this weakness, to the best of our knowledge, this is the first work to leverage temporal reasoning properties as a defense against poisoning attacks in Federated Learning (FL) in general, and Federated Time Series (FTS) specifically. Our approach is pioneering in its use of the semantic behaviors of clients, which presents a novel perspective that is orthogonal to existing model-agnostic defenses in FL. Specifically, we focus on the operational range property and a qualitative semantic evaluation of reasoning logic. Future research could expand upon this by investigating more complex properties, such as nested logic and the quantitative semantics of reasoning specifications. Additionally, there is significant potential to extend this work into other domains where symbolic reasoning-enabled learning can be beneficial, such as healthcare~\cite{antunes2022federated,nguyen2022federated}, where sensitive data must be carefully protected, or autonomous systems~\cite{pola2019control,xianjia2021federated}, where robustness and safety are critical. Since \thename{} is orthogonal to existing defenses, it could be tailored as an add-on component to enhance current defense methods. However, as FL defenses often involve the combination of multiple components, directly integrating \thename{} is not straightforward and would require further effort.
\update{In conclusion, our method shows remarkable improvement in defending against poisoning attacks in FTS. Future works can extend current evaluation setting with more datasets from multiple domains such as Smart Cities, medical analysis and power systems~\cite{krajzewicz2002sumo}~\cite{murat2020application}~\cite{lai2018modeling,zhou2021informer}.}

\clearpage

\end{document}